\documentclass[fleqn,usenatbib]{mnras}

\usepackage{newtxtext,newtxmath}

\usepackage[T1]{fontenc}
\usepackage{ae,aecompl}

\usepackage{graphicx}	
\usepackage{amsmath}	
\usepackage{amssymb}	
\usepackage{enumerate}   
\usepackage{color}
\usepackage{ulem}

\usepackage{lipsum} 
 
\title[The effects of surface fossil magnetic fields on massive star evolution: II]{The effects of surface fossil magnetic fields on massive star evolution: II. Implementation of magnetic braking in MESA and implications for the evolution of surface rotation in OB stars }
 
\author[Z. Keszthelyi et al.]{
Z. Keszthelyi$^{1,2,3}$\thanks{E-mail: z.keszthelyi@uva.nl},
G. Meynet$^{4}$,
M.\,E. Shultz$^{5}$,
A. David-Uraz$^{5}$,
A. ud-Doula$^{6}$,
\newauthor
 R.\,H.\,D. Townsend$^{7}$,
G.\,A. Wade$^{2}$,
C. Georgy$^{4}$,
V. Petit$^{5}$,
S.\,P. Owocki$^{5}$
\\  
$^{1}$Anton Pannekoek Institute for Astronomy, University of Amsterdam, Science Park 904, 1098 XH, Amsterdam, The Netherlands \\
$^{2}$Dept. of Physics and Space Science, Royal Military College of Canada, PO Box 1700, Station Forces, Kingston, ON, K7K~0C6, Canada\\
$^{3}$Dept. of Physics, Engineering Physics and Astronomy, Queen's University, 99 University Avenue, Kingston, ON, K7L~3N6,Canada\\
$^{4}$Geneva Observatory, University of Geneva, Maillettes 51, 1290 Sauverny, Switzerland \\
$^{5}$Dept. of Physics and Astronomy, University of Delaware, 217 Sharp Lab, Newark, DE 19716, USA \\
$^{6}$Dept. of Physics, Penn State Scranton, 120 Ridge View Drive, Dunmore, PA 18512, USA \\ 
$^{7}$Dept. of Astronomy, University of Wisconsin-Madison, 475 N Charter St, Madison, WI 53706, USA \\
}

\date{Accepted XXX. Received YYY; in original form ZZZ}

\pubyear{2020}

\begin{document}
\label{firstpage}
\pagerange{\pageref{firstpage}--\pageref{lastpage}}
\maketitle

\begin{abstract}
The time evolution of angular momentum and surface rotation of massive stars is strongly influenced by fossil magnetic fields via magnetic braking. We present a new module containing a simple, comprehensive implementation of such a field at the surface of a massive star within the Modules for Experiments in Stellar Astrophysics (\textsc{mesa}) software instrument.
We test two limiting scenarios for magnetic braking: distributing the angular momentum loss throughout the star in the first case, and restricting the angular momentum loss to a surface reservoir in the second case.
We perform a systematic investigation of the rotational evolution using a grid of OB star models with surface magnetic fields ($M_\star=5-60$ M$_\odot$, $\Omega/\Omega_{\rm crit} =0.2-1.0$, $B_{\rm p} =1-20$~kG). 
We then employ a representative grid of B-type star models ($M_\star=5, 10, 15$~M$_\odot$, $\Omega/\Omega_{\rm crit} =0.2 , 0.5, 0.8$, $B_{\rm p} = 1, 3 ,10, 30$~kG) to 
compare to the results of a recent self-consistent analysis of the sample of known magnetic B-type stars. We infer that magnetic massive stars arrive at the zero age main sequence with a range of rotation rates, rather than with one common value. In particular, some stars are required to have close-to-critical rotation at the ZAMS.
However, magnetic braking yields surface rotation rates converging to a common low value, making it difficult to infer the initial rotation rates of evolved, slowly-rotating stars. 

\end{abstract}

 


\section{Introduction} \label{sec:intro}

Magnetic fields are routinely detected in stars across the entire Hertzsprung-Russell diagram (HRD), from early to late evolutionary phases \citep{donati2009}. Surface magnetic fields are detected in 7\% of hot, massive, OB stars \nobreak\citep{wade2014,wade2016,martins2015,neiner2015,morel2015,fossati2015,grunhut2017,shultz2018,petit2019}. Unlike those detected in cool stars, these surface fields are likely not being actively generated by a dynamo mechanism, especially because there is no evidence that extended convection zones\footnote{Hot stars do have thin sub-surface layers where inefficient convection (accounting for usually $\approx$ 3\% of the energy transport) occurs due to the iron opacity bump \citep{cantiello2009}, and while there might perhaps be dynamo activity in those layers \citep{cantiello2011}, that would not give rise to the strong, globally organized fields which are observed in magnetic OB stars.} exist at the surfaces of hot stars. The long-term stability and the large-scale structure of these fields (along with the lack of apparent correlation between the field characteristics and stellar or rotational parameters) suggest that these fields are fossil remnants formed during the earlier history of the star \nobreak\citep{cowling1945,mestel1989,moss2003,braithwaite2004,donati2009,neiner2015}. The typical global configuration is approximately a dipole, inclined with respect to the rotation axis, and the polar field strength is usually of the order of a few kG.   

Surface magnetic fields have a complex interaction with stellar winds, confining the wind material along closed magnetic field lines \citep{babel1997,ud2002,owocki2004,townsend2005,bard2016}. This interaction leads to two dynamical effects that have a considerable impact over evolutionary time-scales: \textit{mass-loss quenching}, which reduces the effective mass-loss rate of the star, and \textit{magnetic braking}, which reduces the angular momentum of the star.
 
Early analytic studies described the role of the magnetised solar wind to explain the slow rotation of the Sun \citep{parker1963,mestel1968}. \cite{weber1967} derived a formalism to account for this angular momentum loss, commonly known as magnetic braking. While this term is often used in different contexts, in the following we refer to magnetic braking specifically to describe the rotational spin-down of hot, massive stars caused by a large-scale dipolar surface fossil field. Magnetohydrodynamic (MHD) simulations by \cite{ud2008,ud2009}, specifically applied to OB stars that possess dipolar fields aligned with their rotation axes, have yielded results that are consistent with the analytic formalism derived for a split monopole by \cite{weber1967}.

Observations show that magnetic OB stars rotate more slowly as a population than those OB stars that do not have detected magnetic fields \citep{petit2013,grunhut2017,shultz2018}. Therefore, the spin-down of magnetic stars is expected to be an observable phenomenon. Direct measurements of period change exist for just four magnetic stars: CU Vir, HD\,37776, $\sigma$ Ori E, and HD\,142990. $\sigma$ Ori E's rotation is observed to slow down at approximately the rate predicted by analytical prescriptions of magnetic braking \citep{ud2009,town2010,oksala2012}. Interestingly, in the three other cases, apparently cyclical period changes - including episodes of rotational \textit{acceleration} - have been observed \citep{2011A&A...534L...5M,2019MNRAS.486.5558S}. 

While surface magnetic fields affect the dynamics of the stellar plasma, there has been growing attention toward their long-term evolutionary impact \citep{langer2012}, especially for the following points.
\begin{itemize}
    \item To reconcile the rotation rates and inferred ages of observed magnetic stars using appropriate models \citep{fossati2016,2019MNRAS.490..274S}.
    \item To investigate mass-loss quenching, which was shown to result in an evolutionary channel that may lead to the production of progenitors of heavy stellar-mass black holes and pair instability supernovae even at high metallicity \citep{petit2017,georgy2017}.
    \item To understand the role of magnetic braking in the context of the Hunter diagram, which investigates rotational mixing by showing nitrogen abundance against projected rotational velocity (\citealt{morel2008,hunter2008,hunter2009,brott2011b,meynet2011,potter2012b,martins2012,aerts2014}, and \citealt{keszthelyi2019}, hereafter Paper I).
\end{itemize}  


Surface magnetic braking has been implemented\footnote{see Appendix \ref{sec:genec} for more details} in the Geneva stellar evolution code \citep{eggenberger2008} and in the \textsc{rose} code \citep{potter2012a} in the context of single magnetic OB stars \citep{meynet2011,georgy2017,potter2012b}, and supermassive stars \nobreak\citep{haemmerle2019}. Furthermore, the Modules for Experiments in Stellar Astrophysics (\textsc{mesa}) software instrument \citep{paxton2011,paxton2013,paxton2015,paxton2018,paxton2019} was used to test cases of massive star magnetic braking\footnote{see the summer school material by M. Cantiello: \newline \url{https://doi.org/10.5281/zenodo.2603726}}, and also applied to model the magnetic star $\tau$~Sco \citep{schneider2019}.
Additionally, magnetic braking has also been explored in other contexts, such as binary systems with surface magnetic braking \citep[e.g.,][]{rappaport1983,chen2016,song2018}, and core magnetic braking \citep{maeder2014b,cantiello2016,kissin2018,fuller2019}. Empirical formulae describing surface magnetic braking applicable to low-mass stars (see \citealt{skumanich1972}), specifically in the context of low-mass X-ray binaries, have been studied with the \textsc{mesa} software instrument by \cite{van2019}. Several other studies have also accounted for magnetic braking in low-mass stars (e.g., \citealt{fleming2019}, and references therein). Although previous studies have already used scaling relations to account for magnetic braking in various contexts and evolutionary codes, the implementations of these approaches are not often extensively detailed.

The purpose of this study is to present and elaborate on the implementation of massive star magnetic braking in the open-source software instrument \textsc{mesa} by testing two limiting cases. We developed a module to quantify the impact and time evolution of surface fossil magnetic fields in stellar evolution codes (see also \citealt{keszthelyi2017a}, \citealt{petit2017}, \citealt{georgy2017}, and Paper I). To this extent, we provide a simple implementation of surface fossil magnetic fields in stellar evolution codes. The module is available online on the \textsc{mesa} repository website\footnote{\protect{\url{http://cococubed.asu.edu/mesa\_market/}} \newline \protect{\url{http://doi.org/10.5281/zenodo.3250412}}}.

This work is part of a project in which we aim to systematically explore the effects of surface fossil magnetic fields on massive star evolution. In Paper I, we discussed the qualitative impact of mass-loss quenching, magnetic braking, and magnetic field evolution for a typical massive star model. In this work (Paper II), we focus on the rotational and  angular momentum evolution of a grid of models.

This work is structured as follows. In Section 2, we briefly describe the background of the scaling relations quantifying magnetic braking, and in Section 3, we elaborate on the implementation of magnetic braking in \textsc{mesa}. In Section 4, we explore the parameter space with the computed grid of models, and in Section 5, we discuss the implications of surface magnetic fields and rotation on the Hertzsprung-Russell diagram. In Section 6, we compare our models against the observed sample of magnetic B-type stars, and in Section 7, we summarise our findings.

\section{Scaling relations of magnetic braking} \label{sec:two}

In this section we describe a simple physical model of massive star magnetic braking. 
A key result of the MHD simulations by \cite{ud2009} is that the analytical formalism derived by \cite{weber1967} is an appropriate scaling relation for the angular momentum loss of magnetic massive stars. According to their formalism, the rate of angular momentum removed by the stellar wind and the magnetic field ($\mathrm{d} J_{B} /\mathrm{d} t$) is defined as:
\begin{equation}
\frac{\mathrm{d} J_{B} }{\mathrm{d} t} = \frac{2}{3} \dot{M}_{B=0} \, \Omega_\star \, R_A^2 \, , \label{eq:eq1}
\end{equation}
where $\dot{M}_{B=0}$ is the mass-loss rate the star would have in the absence of a magnetic field (i.e., the wind-feeding rate), $\Omega_\star$ is the surface angular velocity, and $R_A$ is the Alfv\'en radius of the star. A major difference between the works of \cite{ud2002} and \cite{weber1967} is the calculation of the Alfv\'en radius. For a dipolar field configuration, \protect{\cite{ud2002}} introduced the equatorial magnetic wind confinement parameter as
\begin{equation}
\eta_\star = \frac{B_{\rm p}^2 R_\star^2}{4 \dot{M}_{B=0}v_\infty} \, ,
\end{equation}
with $B_{\rm p}$ the polar magnetic field strength, $R_\star$ the stellar radius, and $v_\infty$ the terminal wind velocity\footnote{These quantities are in units of the cgs system.}. 
$\eta_\star$ quantifies the ratio of magnetic to wind kinetic energy.
In terms of $\eta_\star$, the Alfv\'en radius $R_A$ is expressed as:
\begin{equation}
\frac{R_A}{R_\star} \sim 0.29 + (\eta_\star + 0.25)^{0.25} \, , \label{eq:alf}
\end{equation}
for a dipolar magnetic field configuration which is aligned with the rotation axis. In practice, most OB star surface magnetic fields are well described by a dipolar configuration \citep[e.g.,][]{shultz2018}, however, even in the case of more complex field topologies, the dipolar component dominates the angular momentum loss.
%
%
%

The Kepler co-rotation radius, that is the distance at which the centrifugal and gravitational forces are equal to each other, is defined as: 
\begin{equation}
\frac{R_K}{R_\star} = \left( \frac{v_{\rm rot}}{v_{\rm orb}} \right)^{-2/3} =\left( \frac{v_{\rm rot}}{\sqrt{G M_\star / R_\star}} \right)^{-2/3} \, .  \label{eq:rk}
\end{equation} 
where $G$ is the gravitational constant and $M_\star$ is the mass of the star. The phenomenology of the confined wind material depends on the rotation rate of the star. In the case of a slow rotator ($R_A < R_K$) the wind launched from both magnetic hemispheres is trapped and channelled along closed field loops, forming a \textit{dynamical magnetosphere}, and shocks close to the magnetic equator before falling back onto the stellar surface (e.g., \citealt{babel1997,ud2002,owocki2016}). Fast rotators ($R_A > R_K$) additionally form a \textit{centrifugal magnetosphere} as plasma is supported against gravity past the co-rotation radius and accumulates to form dense clouds \citep{townsend2005}. Both types of magnetospheres can be diagnosed using observations in H$\alpha$ \citep{petit2013,shultz2016}.

%
%
%
%
%
%
%
\section{Implementation of massive star magnetic braking in \textsc{mesa}}

\textsc{mesa} \citep{paxton2011,paxton2013,paxton2015,paxton2018,paxton2019} is a rapidly developing, versatile, open-source, one-dimensional stellar evolution software instrument, which provides a flexible way to implement new routines thanks to its modular structure.

The \texttt{run\_star\_extras} module contains a `hook' to implement a desired \texttt{other\_torque} routine. However, the implementation of magnetic braking in stellar evolution codes is not straightforward. This is because one has to define how the angular momentum loss is distributed in the layers of the star. 

We implemented Equation \ref{eq:eq1} and the corresponding equations in \textsc{mesa} version r9793 (and also in later versions, r11701, r12115), and in this work we test two limiting cases of distributing the angular momentum loss. We should note that these two cases, described in detail below, are only interesting so long as solid-body rotation is not enforced. With perfect solid-body rotation (which we cannot fully justify hence the adaptation of \textsc{mesa}'s standard diffusive scheme for rotation) it is always the total angular momentum reservoir that loses angular momentum. 

%
%
\subsection{Key model assumptions}\label{sec:assum}

The key model assumptions are the following:
\begin{enumerate}
    \item The evolutionary models are one dimensional. As a consequence, geometrical effects, such as co-latitudinal variations in the magnetic torque are neglected, the tilt of magnetic fields are ignored, and the variation of mass loss as a function of co-latitude is not taken into account.
    \item Models are not enforced to rotate as solid bodies. In particular, because the interaction between meridional currents and large-scale magnetic fields remains an open question \citep{Maeder2009a}, we did not modify the angular momentum transport to account for Poynting stresses, instead we only considered the losses. Hence, \textsc{mesa}'s diffusive scheme is used with its standard values to account for internal angular momentum transport. This approach allows for testing the impact of angular momentum loss alone.
    \item The interplay between rotation, convection and a fossil field in the stellar interiors is neglected. 
    \item The magnetic torque is assumed to remove angular momentum only from either the near-surface layers or from the entire star. The penetration depth of the fossil field is not assumed to change significantly during the evolution and it is not assumed to depend on the surface field strength.
    \item The magnetic torque is scaled uniformly, thus the fractional specific angular momentum loss remains constant in the considered layers.
    \item The magnetic field evolution model adopts the frozen flux approximation \citep{alfven1942}, therefore the total unsigned magnetic flux through the stellar surface remains constant during the evolution. 
\end{enumerate}

\subsection{The magnetic torque implementation}\label{sec:impl}
The scaling relations from \cite{ud2009} define the rate of angular momentum loss due to wind magnetic braking. However, for a stellar structure model, the rate of specific angular momentum loss d$j_{B}/$d$t$ needs to be considered: 
\begin{equation}
\frac{\mathrm{d}J_{\rm B}}{\mathrm{d}t} = \int_{M_t}^{M_\star} \frac{\mathrm{d}j_{\rm B}  (m)}{\mathrm{d}t} \, \mathrm{d} m = \sum_{k = 1}^{x} \frac{\mathrm{d}j_{\rm B, k}}{\mathrm{d}t}  \, \Delta m_k \, 
\end{equation}
where $m$ is the Lagrangian mass coordinate, $\Delta m_k$ (\texttt{dm\_bar (k)} in \textsc{mesa}) is the mass of a given layer, and $k$ is the index of a layer from the stellar surface inwards\footnote{We will use a notation that is consistent with the \textsc{mesa} instrument papers \citep{paxton2013}.}. The integration limits are defined by the layers of the star where the magnetic torque is operating, that is between an internal layer encompassing a mass $M_t$ (with index $k=x$) and the stellar surface encompassing the total mass $M_\star$ (with index $k=1$). The integration becomes a summation since the mass coordinate takes discrete values in evolutionary codes. The value of $x$ defines the last layer where the magnetic torque is operating. The cumulative mass of layers $k=1...x$ is thus $M_\star - M_t$. The adjustable parameter $x$ is currently a major uncertainty in the model as the penetration depth of the fossil field is unknown\footnote{A further complication that should be elucidated in later approaches is that the penetration depth of the fossil field is likely to be time dependent \citep{braithwaite2006,duez2010}.}.

Using this simple parametric formalism, we investigate two limiting cases of distributing the angular momentum loss in the stellar layers, with the simplifying assumption of consistently keeping the angular momentum transport processes the same in the models. 

\begin{enumerate}[i)]

%
%
\item Angular momentum is extracted only from the near-surface layers of the star:
\begin{equation}
J_{B}^{\rm (SURF)} = \sum_{k=1}^{x} j_{\rm B, k}  \, \Delta m_{\rm k} \, . \label{eq:jbsurf}
\end{equation}
We consider $x$ to be a near-surface zone, namely the last zone unaffected by mass loss (\texttt{k\_const\_mass} in \textsc{mesa}, see \citealt{paxton2013}). This zone varies in time, depending on the mass loss. Typically, $x$ is of the order of 500, while the total number of zones is around 2000-2500 in the models. The mass of these $\approx$ 500 layers is $<< 1\%$ of the total mass of the star as the mass is not equally distributed in the layers. We will denote this case as `SURF' to abbreviate surface angular momentum loss.

%
%
\item Angular momentum is extracted from the entire star:
\begin{equation}
J_{B}^{\rm (INT)} = \sum_{k=1}^{y} j_{\rm B, k}  \, \Delta m_{\rm k} \, \, . \label{eq:jbint}
\end{equation}
In this case, $y$ is equal to the last layer inwards the star (i.e., the summation goes over all layers of the star). Thus this case is equivalent to removing $J_{B}$ from the total angular momentum reservoir of the star $J_{\rm tot}$. We will label this case as `INT' to abbreviate internal, meaning the propagation of angular momentum loss. The SURF case tends toward the INT case if the number of torqued layers are increased (see Appendix~\ref{sec:a1}).
\end{enumerate}

\subsection{The magnetic torque scaling}\label{sec:3.3}

The additional rate of change of the specific angular momentum  can be implemented in the stellar layers as:  
\begin{equation}\label{eq:scal}
\frac{\mathrm{d} j_{\rm B,k} }{\mathrm{d} t} = - \, \frac{J_{\rm B}^{\rm (SURF/INT)}}{J_{\rm SURF/INT}} \, \frac{\mathrm{d} j_{\rm k}}{\mathrm{d} t} \, , 
\end{equation}
where $\frac{\mathrm{d} j_{\rm B,k} }{\mathrm{d} t}$ is the `input' quantity (dubbed as `\texttt{extra\_jdot}' in \textsc{mesa}), d$t$ is one time-step, and $j_{\rm k}$ is the specific angular momentum of a layer with index $k$. Therefore we account for magnetic braking by scaling the specific angular momentum in a given stellar layer according to the ratio of total angular momentum lost (at a given time-step) divided by the total angular momentum budget of the considered layers $\left(\frac{J_{\rm B}^{\rm (SURF/INT)}}{J_{\rm SURF/INT}} \right)$. We assume that a uniformly scaled torque is appropriate thus the fractional specific angular momentum loss remains the same from layer to layer. 

We emphasize that this method does not assume uniform rotation (and does not lead to uniform rotation); instead it \textit{uniformly} decreases the specific angular momentum in the considered stellar layers. Since the rate of change in the additional specific angular momentum must be negative to exert a torque, the above equation appears with a negative sign\footnote{In the previous equations we did not introduce negative signs and the mass-loss rate in Equation~\ref{eq:eq1} appears with a positive sign.}. When magnetic braking is considered for the surface layers, $J_{\rm SURF} = \sum_{k=1}^{x}j_{\rm k} \Delta m_{\rm k}$ is used in the above equation with $x = $ \texttt{k\_const\_mass}. When magnetic braking is considered for all internal layers of the  star, $J_{\rm INT} = \sum_{k=1}^{y}j_{\rm k} \Delta m_{\rm k}$ is used with $y$ being the last layer of the star.

\subsection{Interpretation of the two approaches}

Physically, the loss of angular momentum is always driven from the stellar surface. The simple parameterization we presented allows for testing the case of an assumed coupling (which may or may not be due to the fossil field) that distributes this loss to defined stellar layers. These layers can be adjusted due to empirical constraints.

The SURF method assumes that magnetic braking only taps angular momentum from a limited near-surface reservoir of the star. This case is thus representative of the assumption that the fossil field only penetrates to the near-surface layers and does not directly influence deeper stellar layers. This approach requires a careful consideration since if the layer immediately underneath $x$ is not undergoing an additional specific angular momentum loss per unit time, then the stellar model will begin to enhance angular momentum transport mechanisms to mitigate the resulting loss of specific angular momentum in the layer above. In other words, strong shears will develop if two adjacent layers begin to rotate with significantly different angular velocities.

The INT method may be interpreted as mimicking an efficient coupling between the core and the envelope that propagates the surface angular momentum loss\footnote{We note that while solid-body rotation is not enforced \textit{per se}, in practice the rotation profile of the INT models is nearly flat during the magnetic braking time-scale. This is because angular momentum transport via meridional currents remains efficient in the model.}. Therefore this scenario is representative of the assumption that the fossil field is present throughout the entire star and while it brakes the rotation of the surface layer, it also propagates and distributes the loss to all layers. Alternatively, however, this case may also be interpreted as if the fossil field is not anchored deeply inside the star, but another transport process efficiently propagates the angular momentum loss.

%
%
%

\subsection{Time-step control}

We emphasize that we do not allow for specific angular momentum loss to completely exhaust or exceed the reservoir in a given layer. Nominally, in both (SURF/INT) cases the rate of total angular momentum loss is the same when applying Equation~\ref{eq:eq1}. However, when distributing this loss to a smaller reservoir (SURF case), then it may indeed become comparable to the total angular momentum budget of the considered layers. In this case, decreasing the time-step d$t$ ensures that $J_{\rm B}^{(\rm SURF)}$ cannot approach ${J_{\rm SURF}}$. This is because the former quantity is obtained from Equation~\ref{eq:eq1} as $J_{\rm B}^{(\rm SURF)} = \frac{\mathrm{d}J_{\rm B}}{\mathrm{d}t} \cdot \mathrm{d}t$. In other words, in a shorter time, a smaller amount of angular momentum can be lost.
Because of this, at a given time, the angular momentum lost is not the same in the two (SURF/INT) models. 

This method does not interfere with the calculations in the INT case, however it significantly reduces the time-steps in the SURF case, making it rather computationally expensive. (A time-step may become of the order of 10~yr.) Therefore, a question that we should seek to answer is whether the two methods produce comparable results over evolutionary time-scales. Furthermore, in the stellar evolution model both the losses and the redistribution of angular momentum take place in one time-step; therefore choosing the appropriate time-step alters the strength of these processes (see also, e.g., \citealt{lau2014}).

\subsection{Surface angular velocity}\label{sec:omega}
A key problem is to define the time evolution of the surface angular velocity which is influenced in multiple ways\footnote{We only consider single stars, but evidently the situation is different in binary or multiple systems.}, both by losses and replenishing mechanisms. 

\begin{enumerate}
\item First, even in the absence of magnetic fields, the mass loss due to stellar winds implies angular momentum loss \citep{langer1986, maeder2000, vink2010,rieutord2014,keszthelyi2017b,gagnier2019a}. Therefore, the mass that the star loses via winds also carries away angular momentum, hence decreasing the surface angular momentum reservoir. 

\item Second, in the presence of magnetic fields, angular momentum is removed due to the capability of electromagnetic fields to transport energy and (angular) momentum via Poynting stresses. The angular momentum loss by magnetic fields generally supersedes the angular momentum loss by the stellar winds by an order of magnitude \citep{ud2008}. Since the prescription of \citet{ud2008,ud2009} contains the gas-driven angular momentum loss, it accounts for angular momentum loss even in the absence of magnetic fields, i.e., d$J_B/$d$t \neq 0$ when $B_{\rm p} = 0$.\footnote{This means that when implementing the prescription of magnetic braking (Equation \ref{eq:eq1}), the mass-loss driven angular momentum loss needs to be subtracted to avoid double counting.} 

\item Stellar rotation induces various instabilities in stellar interiors. Although approximate prescriptions have been developed to take these into account in one-dimensional models \citep[e.g.,][]{endal1978,pin1989,zahn92,maeder1998,heger2000,meynet2013}, the rigorous validation of transport processes in massive stars is still underway. Although the pulsational properties of O-type stars are mostly unknown, in principle, asteroseismology is a powerful tool to gain such information and has indeed been successfully applied in the case of a small sample of pulsating B-type stars
(e.g., \citealt{briquet2012,moravveji2015,papics2017,bram2018,handler2019}). 

In fact, presently-used prescriptions for the angular momentum transport in stars are challenged by  asteroseismic constraints (see \citealt{cantiello2014,maeder2014b} and recently \citealt{aerts2019}), showing that the presently-used prescriptions are not efficient enough to reproduce the contrast between the rotation of the core and the rotation of the envelope in sub giants and in red giant stars\footnote{We caution that these results are based on a sample of low and intermediate-mass stars below 8\,M$_\odot$, therefore the conclusions may differ for massive stars. Measurements of the rotation profiles of massive stars are still largely lacking.}. \cite{fuller2019} and \cite{ma2019} proposed a modification of the Spruit-Tayler dynamo allowing to obtain a better agreement with the observations; however, according to \cite{eggenberger2019}, some physics is still missing for providing a satisfactory fit of all the observational constraints.

In the present work we do not discuss how the results will change when varying the internal angular momentum transport. At least for the main-sequence phase, incorporating the effects of a more efficient angular momentum transport is not expected to significantly change the results since the \textsc{mesa} models already rotate nearly as a solid body during that phase. 

Nevertheless, the assumptions made regarding the angular momentum redistribution inside the star may alter the strength of surface magnetic braking. Therefore, while Poynting stresses are expected to transport angular momentum inside the star, it is presently unclear how this would affect other instabilities (e.g. meridional currents) and how the internal rotation profile of massive stars could be characterised. This fundamental uncertainty motivates our approach to test only the angular momentum loss in a first step.
\end{enumerate}

\subsection{Slow-rotation limit}

A caveat of the implemented scaling relations is that the use of Equation \ref{eq:eq1} becomes problematic when the surface of the star has almost completely spun down, that is, when the surface angular velocity approaches zero,
\begin{equation}
\lim_{\Omega_\star \to 0} \frac{\mathrm{d} J_{B} }{\mathrm{d} t} = 0 \, .
\end{equation}
This results in numerical noise if the redistribution processes rapidly enhance the surface angular momentum reservoir by extracting and transporting angular momentum from the core. Because physically it is unclear whether or not the star would indeed enhance transport mechanisms to compensate for such a scenario, we did not manipulate or turn off Equation \ref{eq:eq1} for an arbitrary threshold. 

\subsection{Field evolution and mass-loss quenching}
We complement the \texttt{other\_torque} routine used for magnetic braking with an evolving surface magnetic field as accomplished previously by \cite{petit2017}, \cite{georgy2017}, and Paper I. This means that the polar magnetic field strength is obtained from magnetic flux conservation as:
\begin{equation}
B_{\rm p} (t) \, R_\star^2 (t) \, = \, B_{\rm p,0} R_{\star,0}^2  \, ,     
\end{equation}
with $B_{p,0}$ being the initial surface polar magnetic field strength and $R_{\star,0}$ being the initial stellar radius (see also Paper I).

The presence of large-scale magnetic fields leads to a reduction of the effective mass-loss rate of the star \citep{ud2002,owocki2004,bard2016}. Magnetic mass-loss quenching is taken into account via the \texttt{other\_wind} routine (see also \citealt{petit2017}, \citealt{georgy2017}, and Paper I), where the mass-loss rates are systematically scaled according to the escaping wind fraction\footnote{The mass-loss quenching parameter $f_{\rm B}$ was shown to contain a small correction term when models with rotation are considered \protect{\citep{ud2009}}. To the first order, we neglect this in the present study since, for example, the appropriate geometrical correction for oblique rotation and the latitudinal mass-flux dependence may lead to a much more notable impact. Nevertheless, we recommend taking it into account in future approaches.}
\begin{equation}
f_{\rm B} = 1 - \sqrt{1 - \frac{1}{R_c}} \, = \frac{\dot{M}}{\dot{M}_{B=0}} \, , 
\end{equation}
where $R_c$ is the closure radius defining the distance from the stellar surface to the last closed magnetic loop, and following \cite{ud2008} can be obtained from the Alfv\'en radius (Equation \ref{eq:alf}) as:
\begin{equation}
R_c \sim R_\star + 0.7 ( R_A  - R_\star) \, .
\end{equation}
%
%
%
These points are important because they play a role in the angular momentum evolution of the star. In particular, as the star evolves, $B_{\rm p}$ weakens with time as long as the stellar radius increases (which is generally the case on the main sequence). Therefore a constant magnetic field strength (i.e., increasing magnetic flux) overestimates magnetic braking and mass-loss quenching. However, if mass-loss quenching were more effective, it would also help retain not only more mass but more angular momentum. The escaping wind fraction $f_B$ for O and early B stars is of the order of 10-20\% (Paper I).

%
%
%
\begin{table*}
\caption{Grid of \textsc{mesa} models ($Z = 0.014$) in this study and observed Galactic B-star sample from \citet{shultz2018}.}
\centering
\begin{tabular}{llcccc}   
\hline \hline
&Braking scheme & $M_{\rm \star, ini}$ [M$_\odot$]  & $\Omega_{\rm ini}/\Omega_{\rm crit, ini}$ & $B_{\rm p, ini}$ [kG] & number of models\\
& when $B_{\rm p} \neq 0$ &  & &  & \\
\hline \hline
&& 5, 10, 20, 40, 60 & 0.4 & 0, 5 & 15 \\
Parameter test: &INT / SURF & 40 & 0.2, 0.4 , 0.6, 0.8, 1.0 & 0, 5 & 12 \\
&& 40 & 0.4 & 0, 1, 2, 5, 10, 20 & 8 \\
\hline
& & & & & \\
B-star models:  &INT / SURF & 5, 10, 15 & 0.2, 0.5, 0.8 & 1, 3, 10, 30 & 72\\
& & & & & \\
\hline \hline 
&Braking scheme & $M_{\rm \star, current}$ [M$_\odot$]  & $\Omega_{\rm current}/\Omega_{\rm crit, current}$ & $B_{\rm p, current}$ [kG] & number of stars\\
\hline \hline
B-star sample:  & unknown & 4.3 - 17.5 & 8.9$\cdot$10$^{-5}$ - 0.76 & 0.07- 23.06 & 55\\
\end{tabular} \label{tab:t1}
\end{table*}

\subsection{Revision in the mass-loss scheme and rotational enhancement}
In the \texttt{run\_star\_extras} module we adopted updates/revisions following \cite{keszthelyi2017b}.

We calculate the electron scattering opacity to obtain the Eddington parameter for the pure electron scattering case (following the work of \citealt{kudritzki1989}). Then, the escape velocity is obtained as:
\begin{equation}\label{eq:vesc}
v_{\rm esc} = \sqrt{\frac{2 G M_\star}{R_\star} (1 - \Gamma_e)} \, , 
\end{equation}
where $G$ is the gravitational constant, $M_\star$ is the stellar mass, and $\Gamma_e$ is the Eddington parameter for pure electron scattering. From the escape velocity, the terminal wind velocity is calculated such that $v_\infty = 2.6 v_{\rm esc}$ for $T_{\rm eff} > $ 20 kK, and $v_\infty = 1.3 v_{\rm esc}$ for $T_{\rm eff} < $ 20 kK \citep{kudritzki2000,vink2000}. Consistently with the terminal velocity calculation, we also adopt the first bi-stability jump temperature at its revised value at 20 kK \citep{petrov2016,keszthelyi2017b,vink2018}, in contrast to the predictions of around 25 kK by \cite{vink2000}. This might indeed be significant since older evolutionary model calculations overestimate this value by 5 - 7 kK, and produce a large jump in mass loss (thus also in angular momentum loss) at those effective temperatures (see \citealt{keszthelyi2017b}).

Although \textsc{mesa} contains a built-in calculation (following the work of \citealt{friend1986}) for the rotational enhancement of the mass-loss rates $f_{\rm rot}$ (dubbed as `rotational $\dot{M}$ boost', see \citealt{paxton2013}, and also \citealt{keszthelyi2017b}), we adopted an alternative description based on the work of \cite{maeder2000} (their equation 4.30), which takes into account gravity darkening. In principle, the enhancement factor remains small unless the surface rotational velocity is close to its critical value\footnote{We systematically use $\Omega$ to denote the angular velocity (and $\Omega_\star$ to denote the surface angular velocity), although some other conventions prefer $\omega$. Since in the computed models the equatorial and polar radii differ negligibly, we follow the \textsc{mesa} definition to adopt the critical value of the angular velocity as $\Omega_{\rm crit} = \sqrt{(1-\Gamma) \frac{GM_\star}{R_\star^3} }$. This is nonetheless only used as an input for setting the initial rotation rates, but not for our calculations.}, thus it generally has a small impact in evolutionary model calculations (see also \citealt{keszthelyi2017b} and Paper I). In the cases where the surface rotation is close to its critical value, a more complex formalism should be adopted; however since this would only concern the very early evolution of only one of our models (with $\Omega_{\rm ini}/\Omega_{\rm crit, ini} = 1.0$), we refrain from adopting it. It should also be noted that while in the one-dimensional case the rotational enhancement factor increases the mass-loss rates, \cite{mueller2014} calculated two-dimensional wind models in which the rotation factor led to a decrease in the mass-loss rates. Although this is in contrast with other two-dimensional approaches (e.g. \citealt{gagnier2019a,gagnier2019b}), this is why, for example, \cite{higgins2019} do not consider rotational enhancement in 1D models.

\subsection{General model setup}

The general model parameters are as follows. A solar metallicity of $Z = 0.014$ is adopted with the \cite{asplund2009} mixture of metals, and isotopic ratios are from \nobreak\cite{lodders2003}. The mixing efficiency in the convective core is adopted as $\alpha_{\rm MLT} = 1.5$. Exponential overshooting is used above the convective core with $f_{\rm ov} = 0.024$ and $f_0 = 0.006$ \citep{herwig2000,paxton2013}. In non-rotating models this would roughly correspond to extending the convective core size by 15\% of the local pressure scale height. We adopt these values for all models for simplicity, as the dependence of overshooting on stellar mass and magnetic field strength has not been established yet (although, see \citealt{vandenberg2006} and \citealt{castro2014} regarding the mass dependence.) For stars with surface fossil magnetic fields the core overshooting might be suppressed compared to non-magnetic stars \nobreak (see \citealt{briquet2012} and \citealt{petermann2015}). 

The mass-loss scheme is adopted from \cite{vink2000} and \cite{vink2001} including the change in bi-stability jump temperature (see above), and is systematically scaled by the magnetic mass-loss quenching parameter $f_B$ and rotational enhancement $f_{\rm rot}$. A standard \textsc{mesa} model is considered with respect to rotationally-induced instabilities (following the works of \citealt{heger2000} and \citealt{heger2005}) by adopting the diffusion coefficients that arise from dynamical and secular shear instabilities \citep{endal1978,pin1989}, Eddington-Sweet circulation \citep{eddington1925,sweet1950}, Solberg-H\o iland, and Goldreich-Schubert-Fricke instabilities \citep{goldreich1967,fricke1968}. The Spruit-Tayler dynamo \citep{spruit2002,tayler73} is not considered to operate in the computed models. The rotationally-induced instabilities are scaled by $f_c =~0.033$ for chemical mixing, and the composition gradient $\nabla \mu$ is scaled by $f_\mu = 0.1$ \citep{brott2011,paxton2013}. Since we aim to study the effects of angular momentum loss alone, we do not impose changes on the angular momentum transport or redistribution and apply only the above mentioned configuration. The computed models are summarised in Table \ref{tab:t1}.
 
\subsubsection{Parameter test}

For our parameter tests, we consider a massive star model with an initial mass of 40 M$_\odot$, an initial ratio of surface angular velocity to its critical value of $\Omega_{\rm ini}/\Omega_{\rm crit, ini} = 0.4$, and an initial surface polar magnetic field strength of $B_{\rm p, ini} =$~5~kG. These values for initial rotation and magnetic field strength are typical for an O-type star \citep{petit2013}. We investigate the impact of these parameters by varying one and only one of them at a time (see Section \ref{sec:s4}).  

\subsubsection{B star comparison}

We computed a representative grid of stellar evolution models to compare with observations and parameters presented by \cite{shultz2018} (see Section \ref{sec:s5}). This grid of models encompasses a range of initial stellar masses between 5 and 15 M$_\odot$, initial rotational rates of $\Omega_{\rm ini}/\Omega_{\rm crit, ini}$ between 0.2 and 0.8, and initial polar magnetic field strengths between 1 and 30 kG. 

%
%
%
\section{Parameter test} \label{sec:s4}

%
%
%
%
\begin{figure*}
\includegraphics[width=6cm]{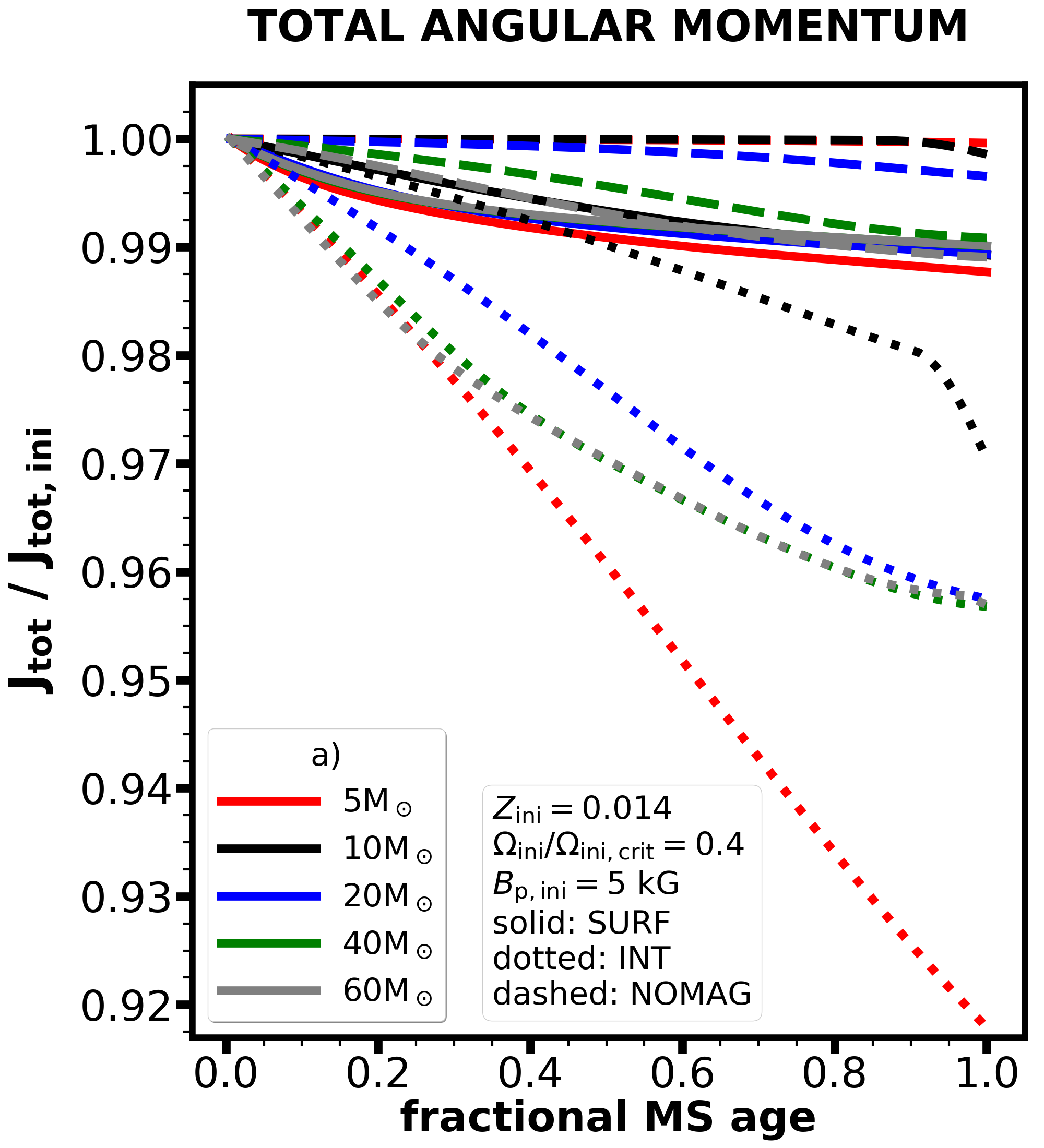}\includegraphics[width=6cm]{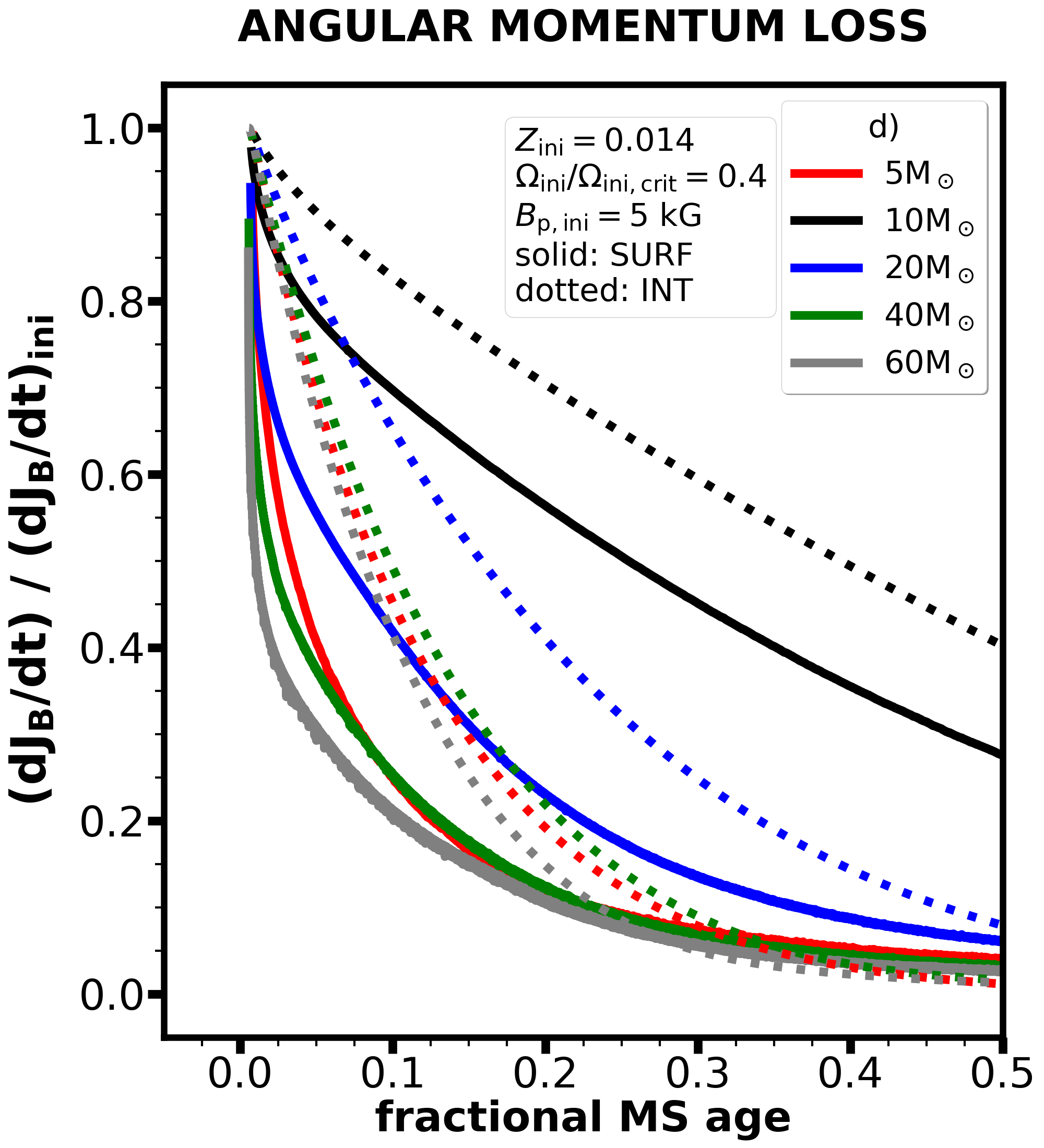}\includegraphics[width=6cm]{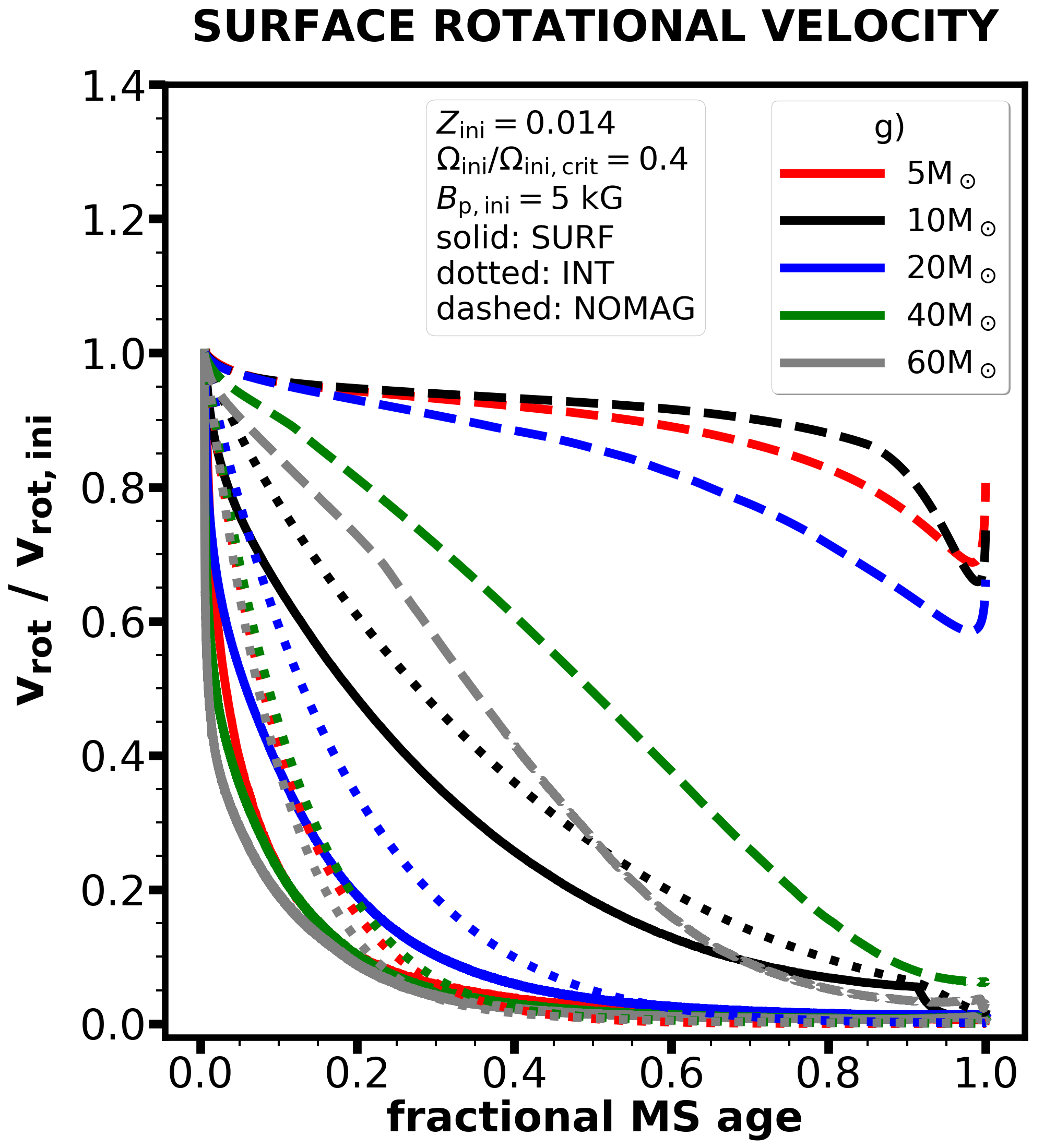}
\includegraphics[width=6cm]{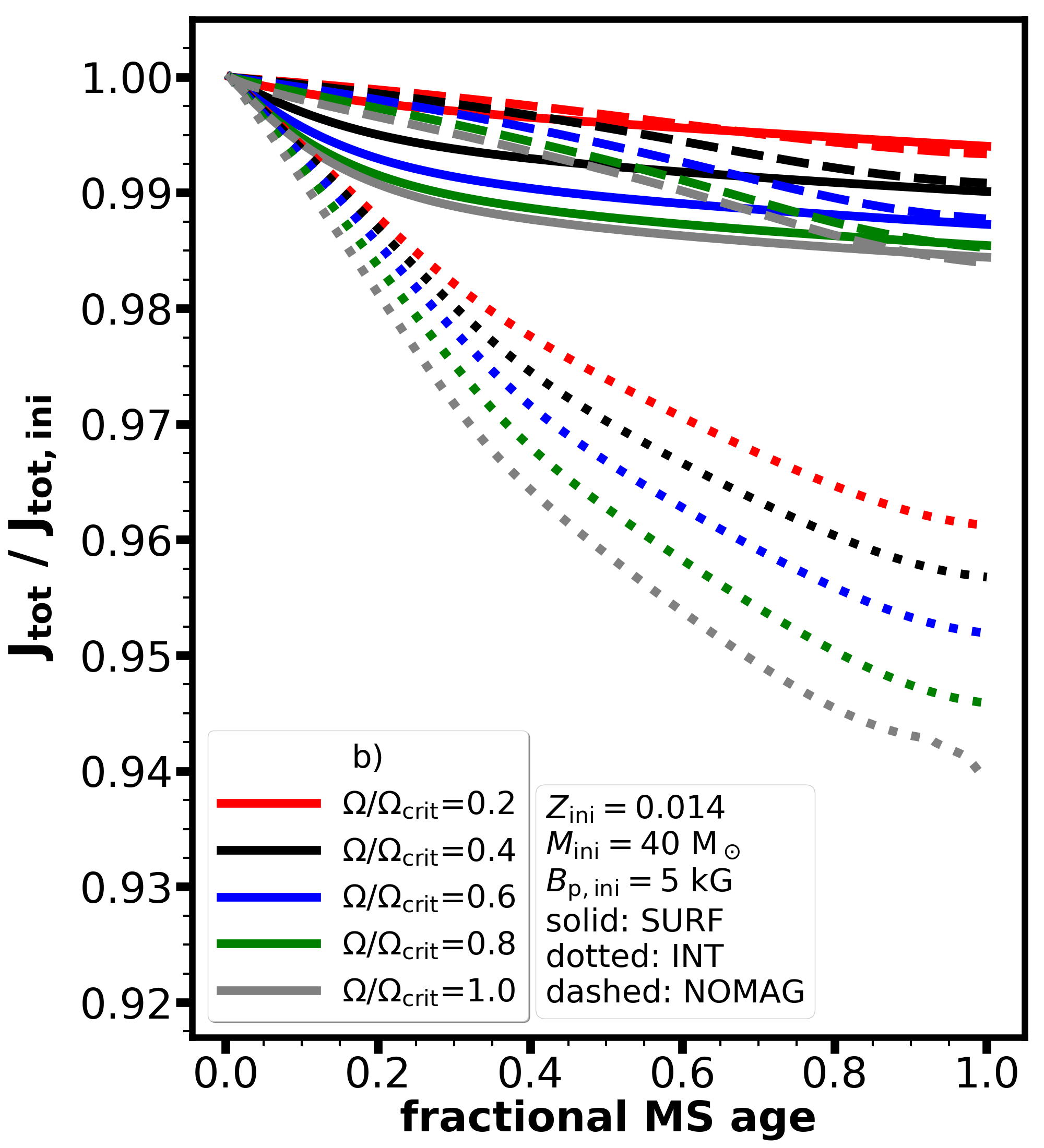}\includegraphics[width=6cm]{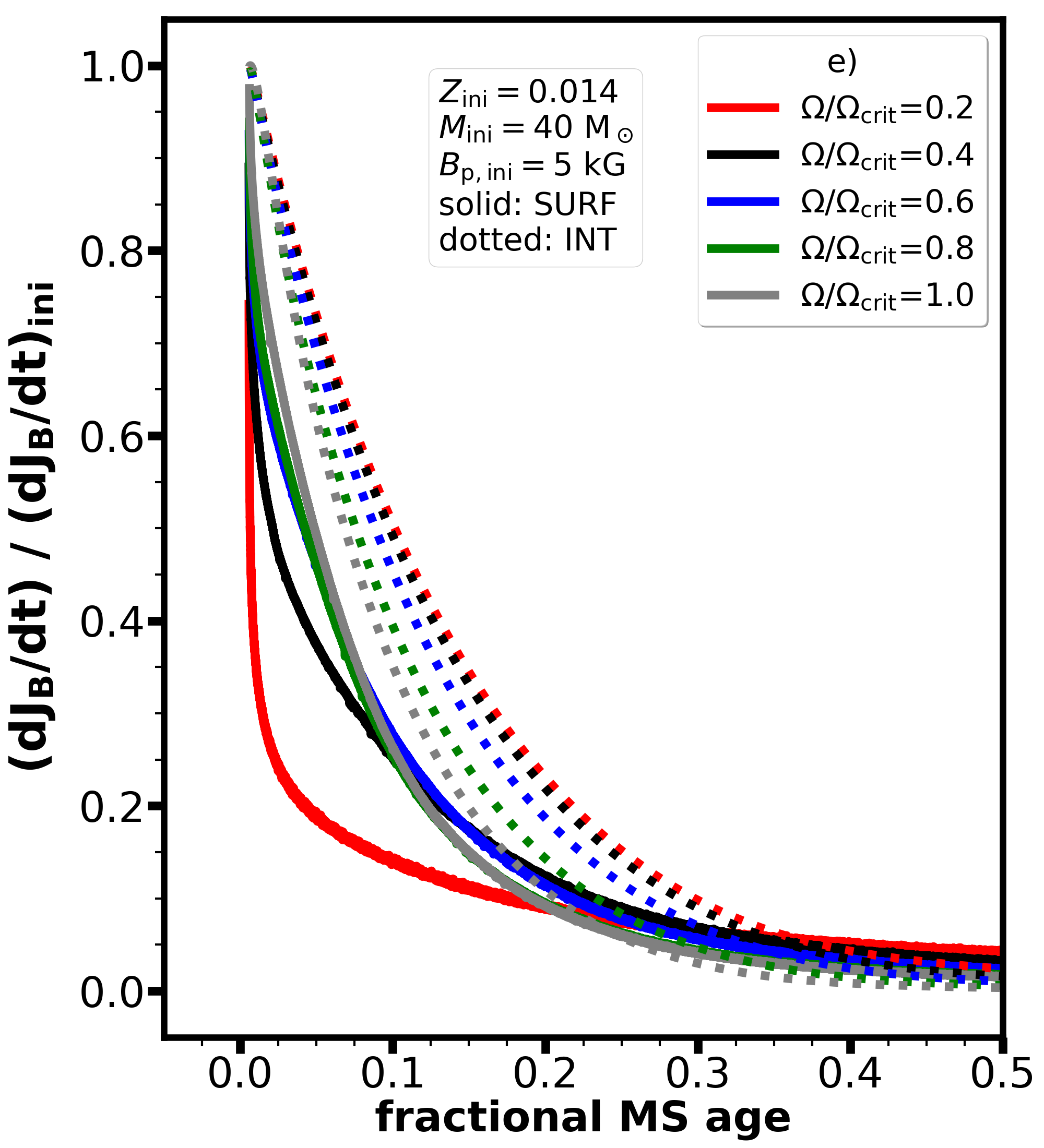}\includegraphics[width=6cm]{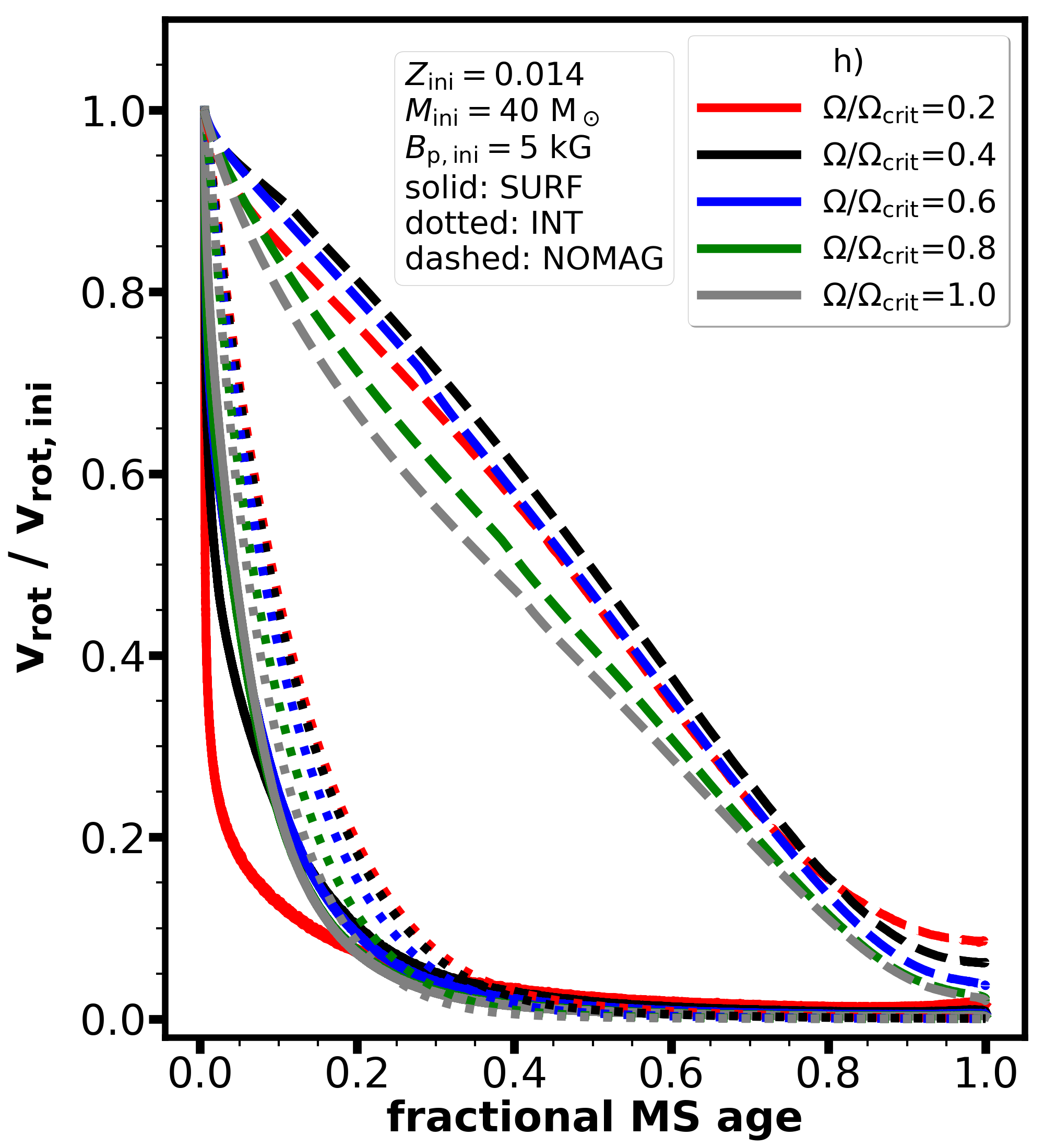}
\includegraphics[width=6cm]{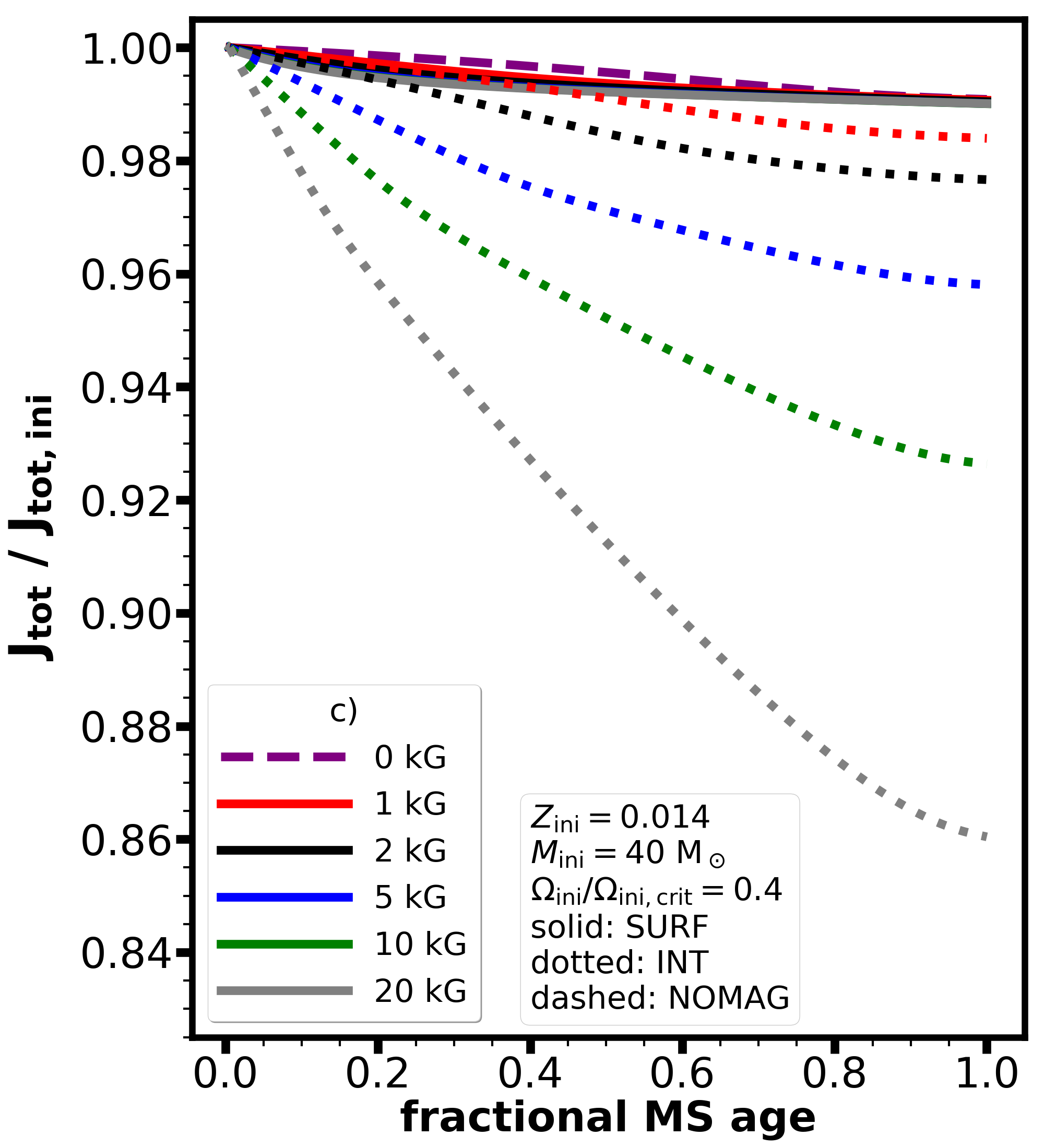}\includegraphics[width=6cm]{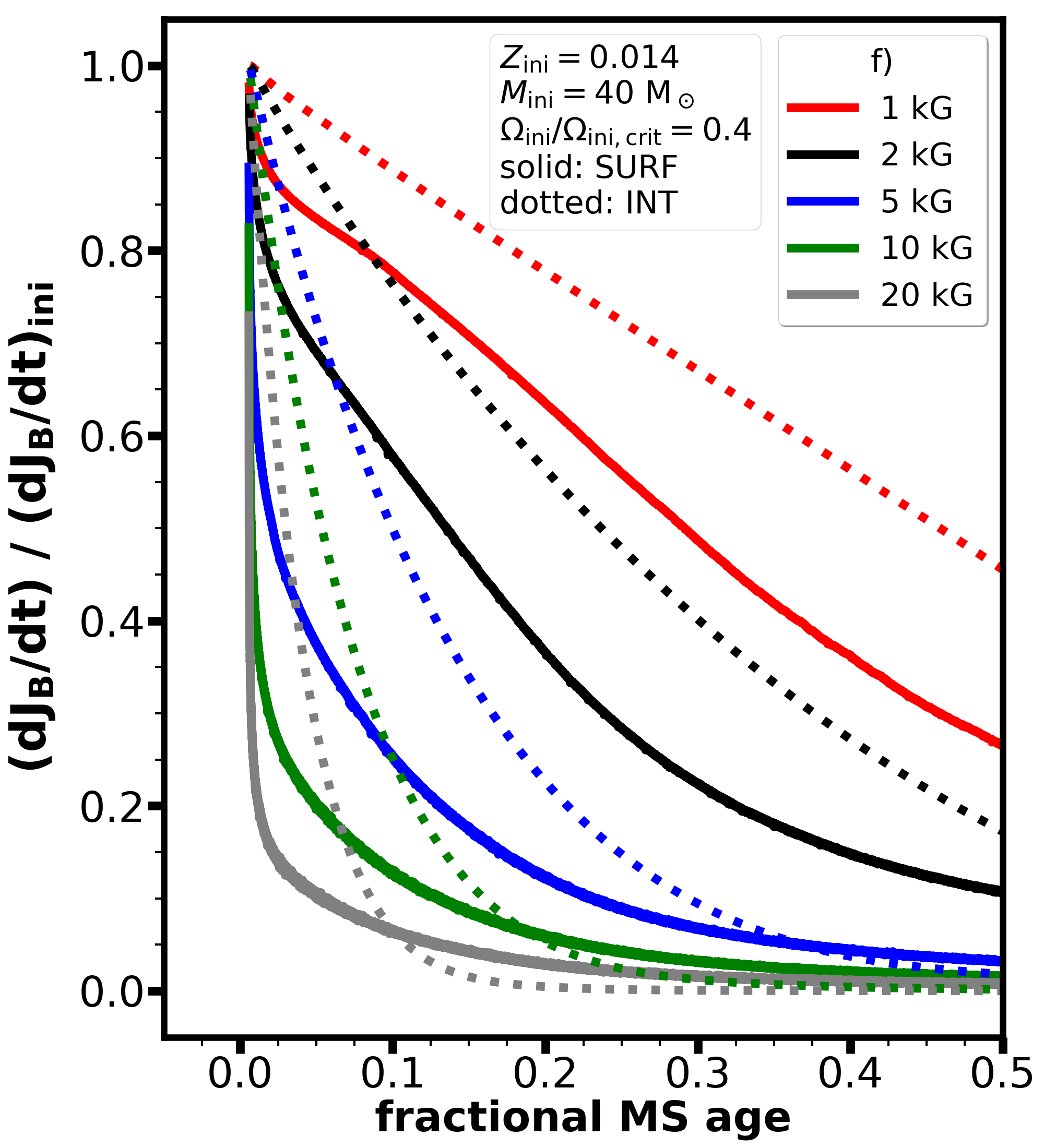}\includegraphics[width=6cm]{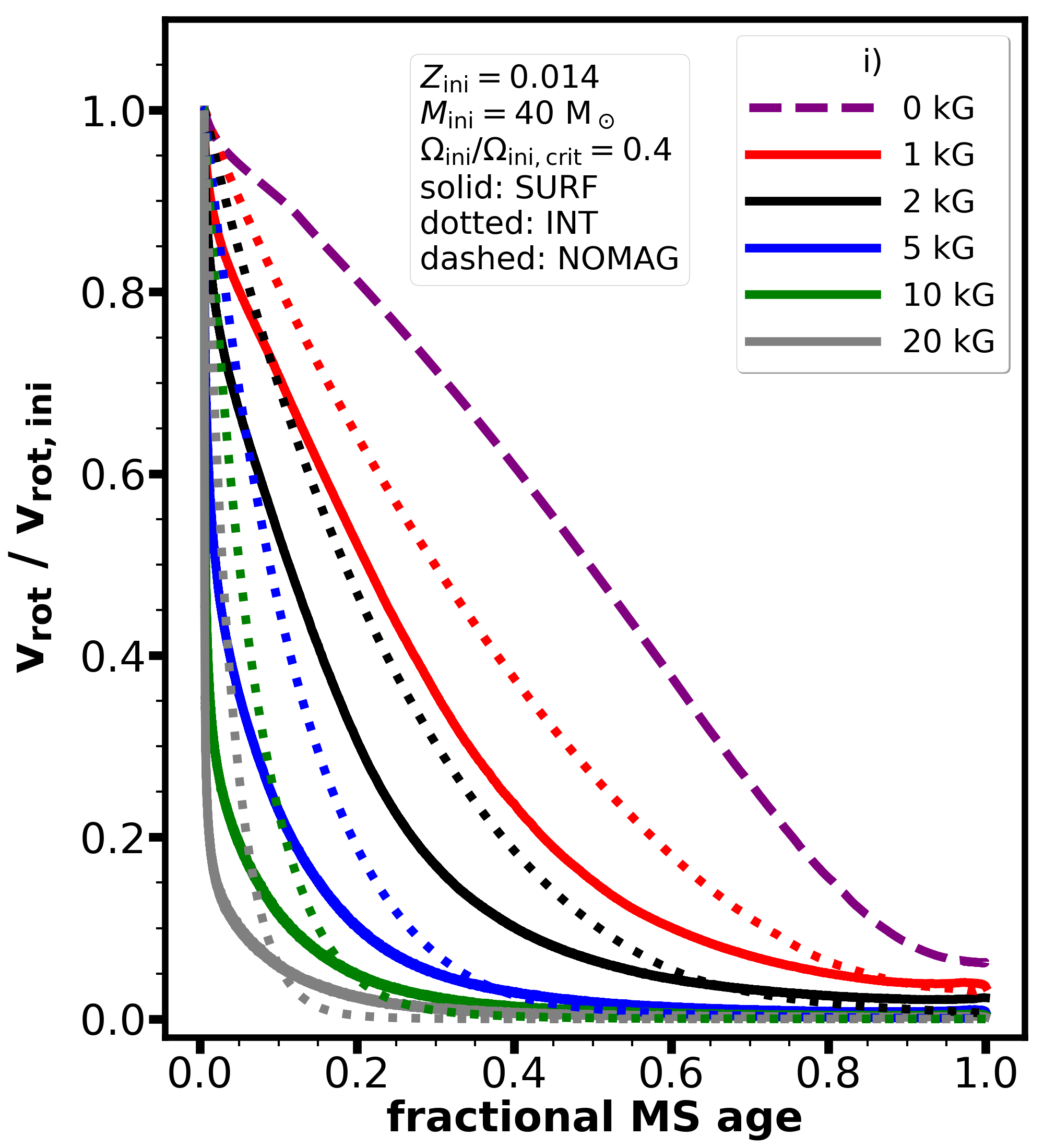}
\caption{Evolution of the total angular momentum (left panels), the angular momentum loss rate driven by the magnetic field and the gas (middle panels), and surface rotational velocity (right panels) compared to their initial values. Varying the mass (top panels), rotation (middle panels), and magnetic field strengths (lower panels) are shown. Models with surface (internal) magnetic braking are shown with solid (dotted) lines. Non-magnetic models are shown with dashed lines. In all models the metallicity is $Z=0.014$, and the same internal angular momentum transport processes are assumed for consistency.}\label{fig:f1}
\end{figure*}

In this Section, we present the results of the parameter tests. Figure \ref{fig:f1} illustrates the effects of wind magnetic braking on the evolution of the total angular momentum (left column), on the rate of angular momentum loss (middle column) and on the surface rotational velocity (right column), in all cases normalised to their initial values. The dependence on initial mass, rotational velocity, and magnetic field strength are shown as top, middle, and bottom panels, respectively. The `default' model is the one with $M_{\star, \rm ini} = 40$~M$_\odot$, $\Omega_{\rm ini}/\Omega_{\rm crit, ini} = 0.4$, and $B_{\rm p, ini} = 5$~kG at solar metallicity ($Z = 0.014$). The solid lines indicate models with surface magnetic braking (SURF), the dotted lines indicate models with internal magnetic braking (INT), and the dashed lines indicate non-magnetic models (NOMAG).

%

\subsection{Evolution of total angular momentum}\label{sec:4.1.1}


%
%
%
%
\subsubsection{Varying initial mass - panel a) }
In the case of non-magnetic models, the angular momentum loss due to stellar winds is stronger for higher-mass models. For a non-magnetic 60\,M$_\odot$ model the main sequence angular momentum loss reaches about 1\% of its initial value. 
When surface magnetic braking is considered, the models lose the same relative amount of angular momentum, around 1\%, independent of their initial mass.
When models account for internal magnetic braking, a larger amount of angular momentum is lost, which is the consequence of more efficiently distributing the losses and also more efficiently replenishing the surface angular angular momentum reservoir, keeping the surface angular velocity higher than in the SURF models. Thus the INT models, which mimic an efficient coupling by removing angular momentum from all layers of the star, always result in losing a larger amount of total angular momentum. 

The 20, 40 and 60\,M$_\odot$ INT models lose the same amount of angular momentum at the end of the main sequence phase (4\,\%)\footnote{The 40 and 60 M$_\odot$ INT models, represented with green and grey dotted lines, practically overlap on the diagram.}, while the 10\,M$_\odot$ loses less (3\,\%), and the 5\,M$_\odot$ model loses about two times as much as the former ones (8\,\%).  

Although the interesting behaviour of the 5\,M$_\odot$ model is amplified for the models with magnetic braking, this trend is also seen for the non-magnetic models. The reason why the lower-mass model loses more angular momentum and brakes its surface rotation more rapidly is a consequence of various factors having counteracting effects. 

Weaker winds (as is the case for lower initial mass) implies weaker angular momentum loss. On the other hand, the quantity of angular momentum that is lost at the surface depends also on the amount of angular momentum flowing from the inner regions of the star to the surface. The more rapid this transport, the larger the amount of angular momentum loss at the surface. The time-scale for this transport varies approximately as the radius of the star divided by the typical value for the meridional velocity in the stellar envelope. When the initial mass of the star decreases, both the radius and the meridional velocity decreases, so that it is difficult to predict what the result will be. Another factor is of course the duration of the main-sequence phase, the more extended it is, the larger the amount of angular momentum loss. The fact that the 5\,M$_\odot$ loses more angular momentum at the end of the MS phase than the 10\,M$_\odot$ is a consequence of all these factors combined.

%
%
%
%
\subsubsection{Varying initial rotation - panel b) }


The non-magnetic and SURF models lose the same amount of angular momentum, while the INT ones lose larger amounts than those models. Nevertheless, in the former two cases, the evolution of the total angular momentum differs during the main sequence. The magnetic SURF models lose a larger amount of angular momentum than non-magnetic models at the beginning of the evolution. This is expected since during the early phases the star expands rather slowly and thus the surface magnetic field does not weaken fast. In the later phases, the SURF models lose less angular momentum because the surface magnetic field strength has decreased and the surface rotational velocity becomes low. The non-magnetic models, in contrast, lose more angular momentum at the end of the main sequence phase, since at that stage they still have  significant surface rotation and the mass loss rates are higher.

%
%
%
%
\subsubsection{Varying initial magnetic field strength - panel c) }


In this parameter test too, surface magnetic braking yields comparable total angular momentum loss as the non-magnetic model. This indicates that surface magnetic braking may only play a minor role in the total angular momentum evolution of the star. The SURF models initially lose more angular momentum, whereas the non-magnetic  models loses more angular momentum at the end of their main sequence phase. The angular momentum evolution of the INT models - in contrast to the other two types of models - show a significant dependence on the initial surface magnetic field strength: the stronger the field, the larger the angular momentum loss.
 
%
%
\subsection{Evolution of magnetic braking}\label{sec:4.2}


%
%
%
%
\subsubsection{Varying initial mass - panel d) }
The change of the rate of angular momentum loss is primarily attributed to the change in three parameters: i) the increase of $\dot{M}$, ii) the decline of $B_{\rm p}$, and the decline of $\Omega_\star$ over time. In a sense, this is a loop of consequences because as magnetic braking decreases the surface angular velocity, the strength of magnetic braking weakens as well (Equation~\ref{eq:eq1}). 

Above 10\,M$_\odot$, there is a clear dependence with stellar mass: magnetic braking weakens faster with higher stellar mass. In the case of the 5\,M$_\odot$ model the fast decrease of magnetic braking is due to fast decrease in the surface rotation rates (panel g) of Figure \ref{fig:f1}). 

%
%
%
%
\subsubsection{Varying initial rotation - panel e) }

The differences between the INT and SURF models decrease when the initial rotation is higher. This is expected since in the SURF models increasing the surface rotation also increases the efficiency of the internal angular momentum transport by the meridional currents and thus these models are approaching the situation realised in the INT models. While the SURF models show a complex behaviour, the INT models reveal that for higher rotation rates the weakening of magnetic braking is more rapid. 

%
%
%
%
\subsubsection{Varying initial magnetic field strength - panel f) }
Varying the magnetic field strength shows that magnetic braking systematically weakens faster with stronger fields. Indeed, over a short time-scale strong fields lead to slowly-rotating models, which will then have weak magnetic braking.

An essential component of the model calculations is to account for the time evolution of magnetic braking. This is in contrast with previous simplifying assumptions, which often extrapolated from current to ZAMS conditions, supposing a constant value of $\frac{\mathrm{d}J_{B}}{\mathrm{d}t}$ over time (e.g. \citealt{petit2013}). Such an assumption fundamentally breaks down and can be especially misleading when the current rotation is slow. We highlight the impact of this issue in Section \ref{sec:spin}.

Strong magnetic braking is not maintained during the entire main sequence evolution (middle panels of Figure~\ref{fig:f1}) because the two necessary ingredients (strong magnetic fields and fast rotation) do not co-exist for considerable time-scales.

\subsection{Evolution of surface rotation} 


%
%
%
%
\subsubsection{Varying initial mass - panel g) }
The non-magnetic 5 to 20\,M$_\odot$ models only have a modest change in their surface rotation during the main sequence. If real stars followed such tracks, it would allow for approximating the value of the initial surface rotation just by measuring the actual one. Magnetic braking makes this connection no longer possible. Independent of the initial rotation or the braking method, all 40\,M$_\odot$ models with $\Omega_{\rm ini}/\Omega_{\rm ini,crit} = 0.4$ converge to an extremely slow surface rotation after one fifth of their main sequence life-time, that is about 1 Myr. Importantly, a low value measured presently does not preclude a high rotation in the past.

%
%
%
%
\subsubsection{Varying initial rotation - panel h) }

All the 40\,M$_\odot$ models arrive at the TAMS with a similarly very low surface rotational velocity regardless the initial rotation. 

In the non-magnetic case, the mass loss by stellar winds are sufficient to remove a large amount of angular momentum. As already noted above for the most massive stars, this would not allow to trace back the initial rotation rates from measuring the surface rotation at an advanced stage of the main sequence evolution. However, at a given fractional main sequence age, the non-magnetic models have much higher surface rotational velocities than the magnetic models.

The SURF models are those presenting the most rapid drop off of the surface rotation. Again this is expected since in this case the time-scale for replenishing the outer layers with internal angular momentum is much longer than the magnetic braking time-scale. The rate of change of the rotation period is very different between the SURF and INT models and thus would represent a way to differentiate between these models, provided of course that some other constraints would allow to give information about the initial mass, metallicity, surface magnetic field, and fractional age.

%
%
%
%
\subsubsection{Varying initial magnetic field strength - panel i) }
%

Clearly, the stronger the magnetic field, the more rapidly rotation brakes. At the early evolution, the INT models maintain a higher surface rotation with respect to the SURF models. In the INT models, the whole star is slowed down at the same time. In the SURF models (also in the non-magnetic one), initially only the outer layers are slowed down hence the more rapid decrease of the surface rotation. 

%
%

\subsection{Impact on spin-down age determination}\label{sec:spin}

\begin{figure*}
\includegraphics[width=18cm]{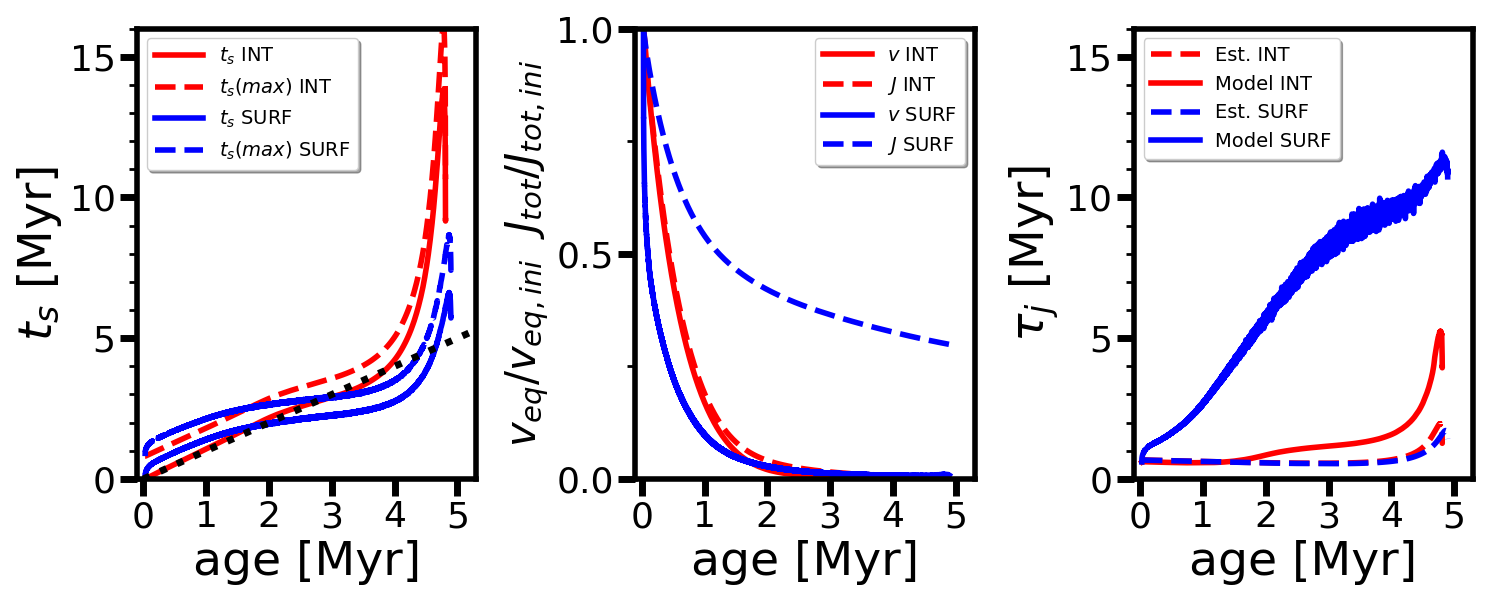}
\caption{Estimated spin-down age (left panel), normalised rotational velocity and total angular momentum (middle panel), and spin-down time-scale (right panel) vs. the age of the modelled star. The two default 40\,M$_\odot$ SURF and INT models are shown as described in Section~\ref{sec:s4}.}\label{fig:app2}
\end{figure*}

The spin-down age derived by \cite{petit2013} relies on two important assumptions:
\begin{itemize}
    \item The rate of angular momentum loss is constant with time.
    \item At any time, the surface rotation is directly tied to the total angular momentum of the star, such that solid-body rotation is achieved. 
\end{itemize}
We have shown that the first assumption is not justified because magnetic braking evolves with time (Section \ref{sec:4.2}), and it remains unknown whether stars could be characterised as solid-body rotators or not.

Assuming the two conditions above, the scaled rotation speed ($W = v
_{\rm rot} / \sqrt{G M_\star / R_\star}$, see also \citealt{ud2008}) can be used to infer the spin-down age such that 
\begin{equation}
t_s = \tau_J \, (\ln W_0 - \ln W)    \, ,
\end{equation}
where $W_0$ is the initial (ZAMS) scaled rotation speed and thus  it is unknown for real stars. Using $W_0 = 1$ for critical rotation yields a maximum spin-down age. Here, we will assume that the initial scaled rotation speed is known and set to the initial rotation of our models. At a given time, the spin-down time-scale is defined by \cite{petit2013} as
\begin{equation}
\tau_J = \frac{J}{\mathrm{d}J/\mathrm{d}t} = \frac{3}{2} f \frac{M_\star}{\dot{M}_{B=0}} \left(\frac{R
_\star}{R_A} \right)^2 \, , 
\end{equation} 
where the first equation can be used to express $\tau_J$ (solid lines, right panel of Figure~\ref{fig:app2}) from the model calculations\footnote{Presently it remains unclear how the spin-down time-scale of the model with surface magnetic braking could be appropriately described. In that case, it is only the surface angular momentum reservoir which is being exhausted not the total angular momentum of the star. However, it is further complicated by the presently unknown angular momentum transport inside the star, which could replenish the surface reservoir with angular momentum from the stellar core.}, whereas the second one can be used to infer $\tau_J$ from observed and estimated quantities (dashed lines on the right panel of Figure~\ref{fig:app2}). Thus as an experiment, to calculate the expected error, $\tau_J$ is expressed knowing $M_\star,\dot{M}_{B=0}, R_\star, R_A $ in a given time, and as \cite{petit2013}, assuming the moment of inertia factor $f=0.1$ throughout. From this estimated spin-down time-scale, we compare the inferred spin-down age (solid lines) to the actual age of the star model on the left panel of Figure \ref{fig:app2}. For these specific models (the two default 40\,M$_\odot$ models described in the beginning of Section~\ref{sec:s4}), the error on the inferred spin-down age is a few Myr initially, while the deviations reach a factor of 3 towards the TAMS. However, the moment of inertia factor is directly proportional to the spin-down age and can introduce severe discrepancies when it is only assumed.

The evolution of the surface rotational velocity (solid lines, middle panel) is closely tied to the evolution of the total angular momentum (dashed lines) in the INT case, whereas for the SURF case, the evolution of the surface rotational velocity and the total angular momentum are decoupled (making the second assumption inappropriate). For the INT model, both the estimated spin-down time-scale and the model spin-down time-scale ($J/\frac{\mathrm{d}J}{\mathrm{d}t}$) are approximately constant over the first few Myr (right panel). This means that the spin-down age only begins to significantly deviate from the real age once the star is close to the TAMS - in this particular setup.

This brief analysis is a best case scenario, in which the stellar parameters are accurately known. It is clear that the model with surface magnetic braking invalidates the original assumptions made to calculate spin-down ages, therefore, further considerations are required. In this case, the spin-down time-scale would need to be defined as
\begin{equation}
    \tau_J = \frac{\Omega_\star}{\mathrm{d}\Omega_\star/\mathrm{d}t} \, ,
\end{equation}
despite that the surface angular velocity is influenced by multiple processes (Section~\ref{sec:omega}), and thus cannot be described by magnetic braking alone.

%
%



%
%
\section{Progenitors of slow rotators}

\begin{figure}
\includegraphics[width=9cm]{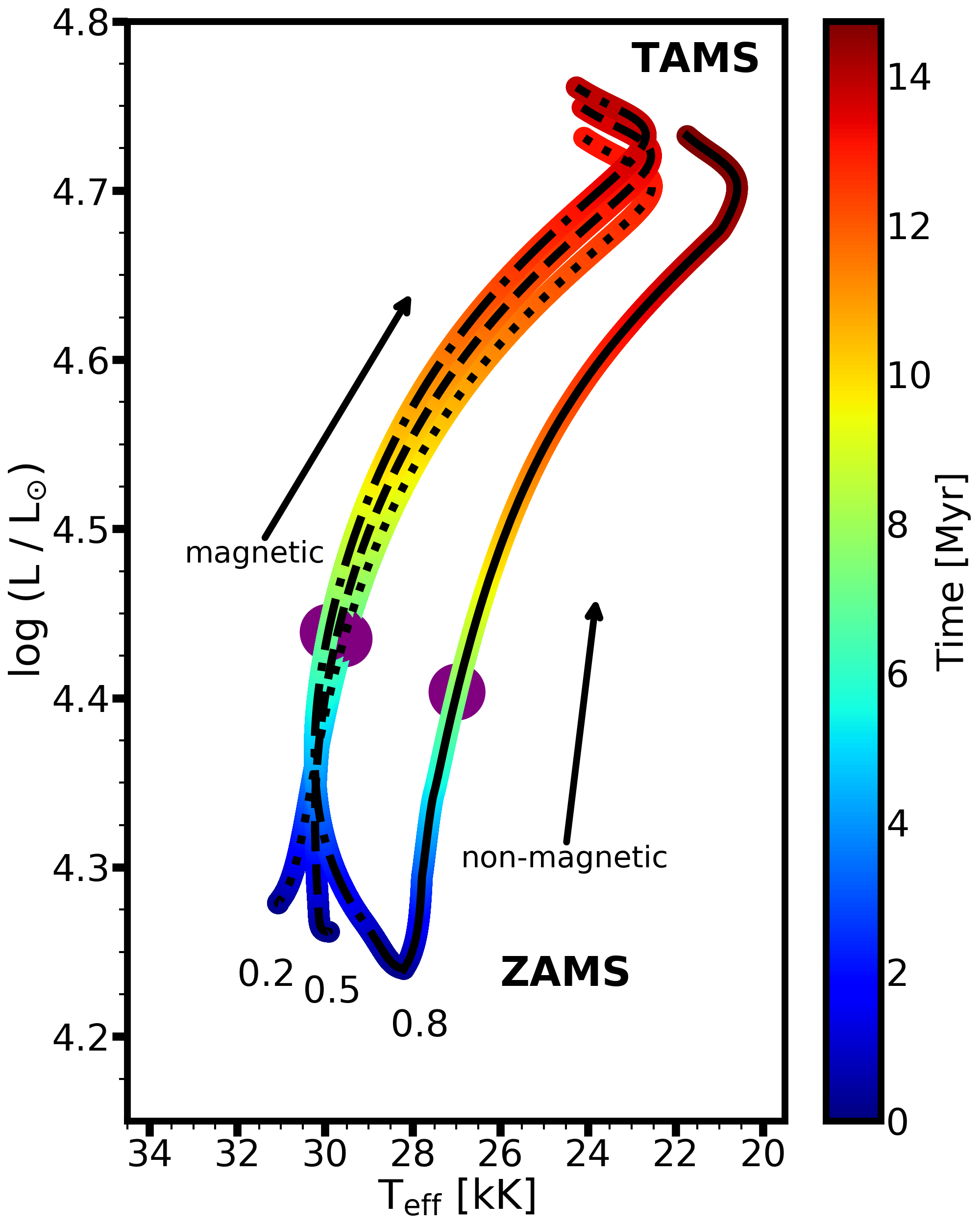}
\caption{Models with different initial ratios of $\Omega/\Omega_{\rm crit}$ (indicated next to the tracks: 0.2 dotted line, 0.5 dashed line, 0.8 dash-dotted line, respectively) are shown in the case of the 15 M$_\odot$ models with an initial polar field strength of 3 kG. A reference non-magnetic model (solid line) is also shown with the same mass and an initial ratio of $\Omega/\Omega_{\rm crit} = 0.8$. The colour-coding indicates the time after ZAMS and the purple markers show the location at which the model is exactly half-way through its main sequence lifetime. Models with surface magnetic braking are shown.} \label{fig:mhh}
\end{figure}

Figure \ref{fig:mhh} shows models of massive stars with initially 15~M$_\odot$. The inclusion of surface fossil magnetic fields results in two notable features on the HRD: i) models with the same initial mass and initial field strength but different initial rotational velocities follow the same track after their early evolution, leading to a degeneracy between them, and ii) the evolutionary track of an initially fast-rotating magnetic model differs significantly from an initially fast-rotating non-magnetic model. 

With higher rotational velocities, the ZAMS position of the tracks are shifted to lower luminosity and effective temperature. This is due to the mechanical effects of rotation: the centrifugal force contributes to balance gravity. The rapidly-rotating model therefore has a ZAMS position that resembles of a non-rotating star with a lower initial mass. Since the wind magnetic braking slows down the surface rotation of the star, the magnetic models converge towards the slow-rotating case independent of the initial rotation.

However, when magnetic braking is absent, the star retains more angular momentum and the track remains in the redder part of the HRD due to the mechanical effects of rotation, that is, the star evolves with lower effective temperature. If the mixing was more efficient, it would result in a blueward evolution on the HRD. For fast-rotating models a blueward evolution is indeed associated with chemically homogeneous evolution (see e.g. Figure 6 of \citealt{brott2011}). Contrary to that, the quantitative differences in the initially rapidly-rotating magnetic model arise from the fast disappearance of the initial centrifugal support, which leads to the tracks converging towards the slow-rotating case.

Finally, if observed stars were attempted to be reconciled with non-magnetic evolutionary models, their physical parameters would only be consistent with initially slowly-rotating models. Instead, observed stars should also be contrasted with models that initiate their evolution as rapid rotators but brake their rotation due to magnetic braking. The physical characteristics of the two models (initially slow and fast rotators, respectively) are different.

For reference, at half-way through their main sequence evolution (shown with purple markers on Figure~\ref{fig:mhh}) the magnetic (SURF) and non-magnetic models with $\Omega_{\rm ini}/\Omega_{\rm ini, crit} = 0.8$ have stellar ages of 6.9 Myr and 7.3 Myr, while at the TAMS their ages are 13.9 Myr and 14.7 Myr, respectively (see Tables~\ref{tab:t2} and \ref{tab:t3}). This may lead to systematic shifts in the estimated stellar ages of observed stars, therefore it has to be considered when comparing observed magnetic massive stars with evolutionary models that do not account for surface magnetic fields \citep{fossati2016,schneider2016,shultz2018}. 

%
\section{Comparison of model predictions with the population of magnetic B-type stars}  \label{sec:s5}
%
%
%
%
%
\cite{shultz2018} presented high-resolution magnetometry and obtained rotation periods for the known population (55) of main sequence magnetic B-type stars with spectral types between B5 and B0. We use this sample of magnetic stars to evaluate our magnetic B-star evolutionary models. Atmospheric parameters for these stars based upon spectroscopic modelling and Gaia data release 2 parallaxes were presented by \cite{shultz2019}. \cite{2019MNRAS.490..274S} combined these observables to calculate fundamental stellar parameters, magnetic parameters (assuming an inclined dipole topology), stellar wind parameters, and magnetospheric parameters. The estimated masses of the observed stars range from 4.3 to 17.5 M$_\odot$, and were determined from the non-magnetic evolutionary models of \cite{ekstroem2012}.

\begin{figure*}
\includegraphics[width=18cm]{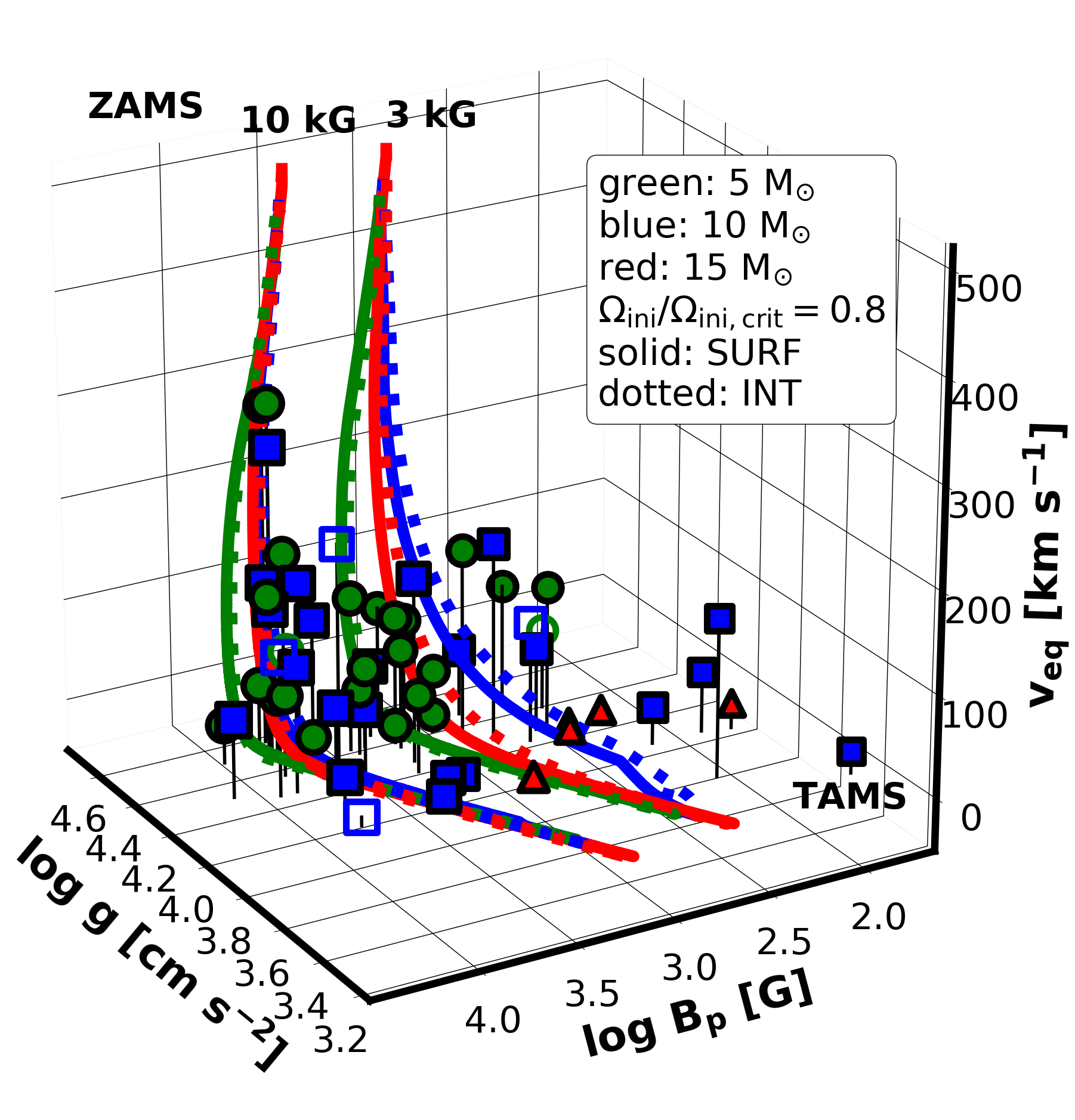}
\caption{Shown are the surface equatorial rotational velocities, the dipolar magnetic field strength and the logarithmic surface gravities of the observed stars and model predictions. The observations are separated into three mass bins: 5 (green circles), 10 (blue squares) and 15 (red triangles) M$_\odot$. Models with $\Omega_{\rm ini}/\Omega_{\rm ini,crit} = 0.8$ and $B_{\rm p, ini} =$ 3 and 10 kG are shown. Known short-period binaries are shown with open symbols.}\label{fig:bstar1}
\end{figure*}

\begin{figure*}
\includegraphics[width=9cm]{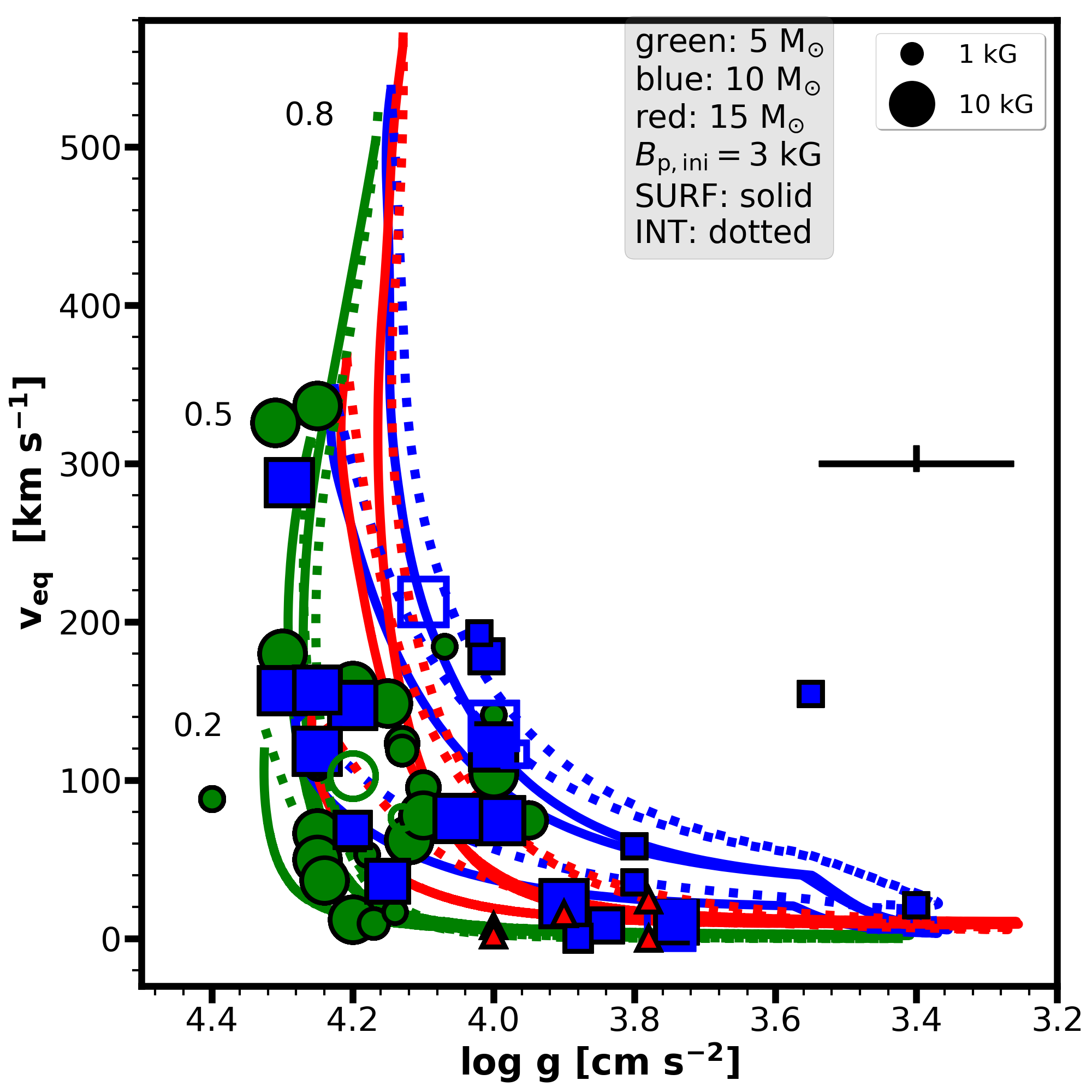}\includegraphics[width=9cm]{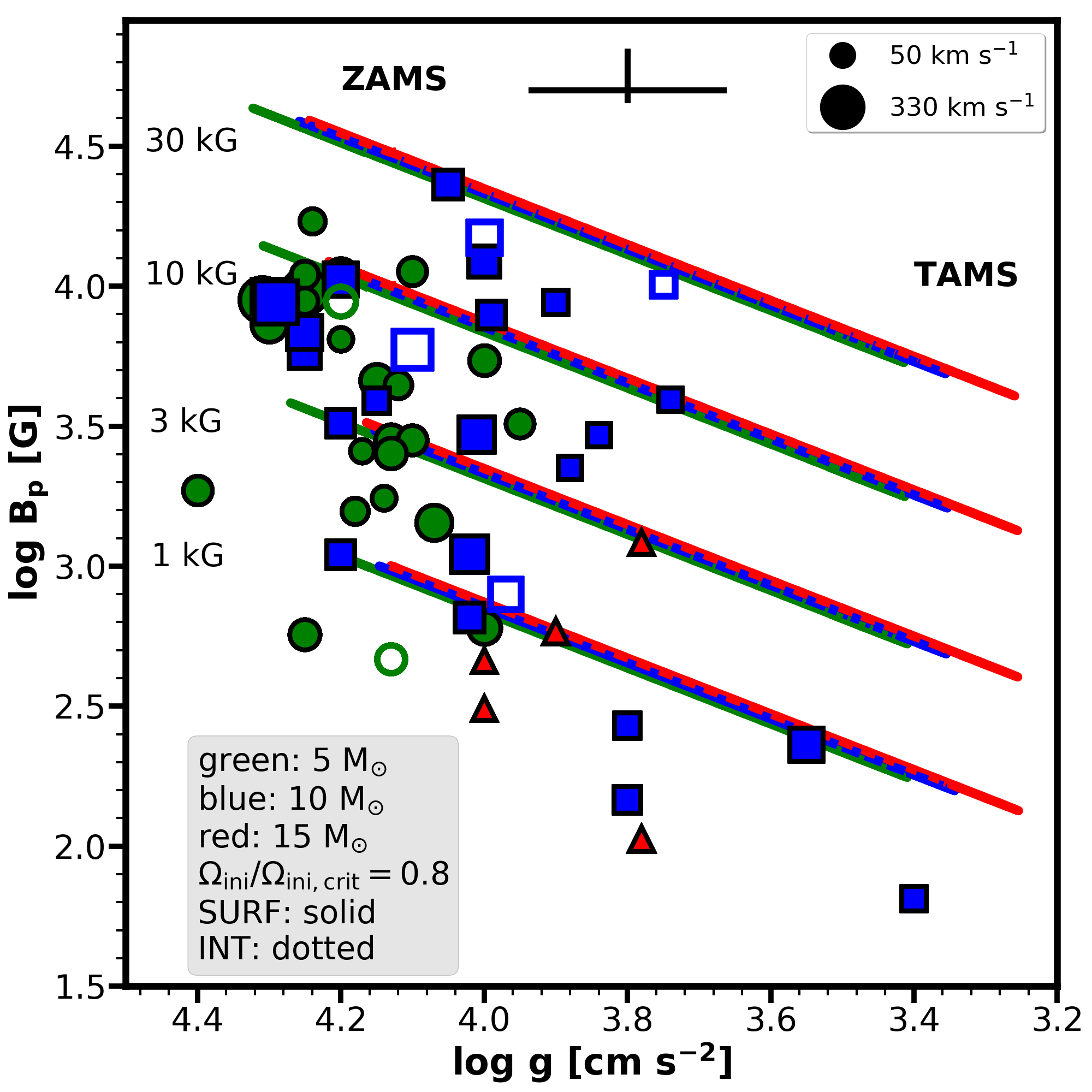}
\caption{Shown are the surface equatorial rotational velocities (left panel) and dipolar magnetic field strengths (right panel) vs. the logarithmic surface gravities. The colours follow the description of Figure~\ref{fig:bstar1}. The mean errorbars are indicated to the centre right and top, respectively. Markers vary in size corresponding to the measured dipolar field strength of observed stars (left panel) or the surface equatorial rotational velocities (right panel). \textit{Left panel:} Models with $\Omega_{\rm ini}/\Omega_{\rm ini,crit} = 0.2, 0.5, 0.8$ and an initial polar magnetic field strength of 3 kG are shown. \textit{Right panel:} Models with $\Omega_{\rm ini}/\Omega_{\rm ini,crit} = 0.8$ and an initial polar magnetic field strength of 1, 3, 10, and 30 kG are shown.
 }\label{fig:bstar2}
\end{figure*}

To facilitate a more straightforward comparison, we separated the observed stars into three mass bins: $M_\star < 7.5\,\mathrm{M}_\odot$, $7.5 < \mathrm{M}_\odot <12.5$, and $\mathrm{M}_\odot > 12.5\,\mathrm{M}_\odot$ (coloured with green, blue, and red in Figures \ref{fig:bstar1}-\ref{fig:bstar4}), and we computed 5, 10, and 15 M$_\odot$ models correspondingly. Known short-period ($P_{\rm orb} < 2~$~days) binary stars (HD\,37017, HD\,149277, HD\,136504, HD\,156324, HD\,36485) are shown with open symbols. 
%
%
%

%
%
%
%
%
%

\subsection{Evolution of surface rotation and magnetic field strength}
 
Figure \ref{fig:bstar1} shows the evolution of surface rotational velocity and magnetic field strength, in terms of the surface gravity. The 3D plot shows models with initially 3 and 10 kG fields, initiating their evolution on the top left and evolving towards the lower right of the diagram. This diagram emphasizes that the models need to account for multiple observables simultaneously. The left panel of Figure \ref{fig:bstar2} further details the rotational evolution, and models with different initial rotation rates are shown for the 3 kG field strength, which was chosen as  \cite{2019MNRAS.490..274S} found a mean field strength close to that value. The right panel of Figure~\ref{fig:bstar2} displays the magnetic field evolution, showing models with varying magnetic field strength, keeping the initial ratio of $\Omega/\Omega_{\rm crit} = 0.8$. On all three diagrams, the colour-coding represents stellar mass (see caption).

\subsubsection{Rotational evolution of known magnetic B-type stars}


From Figures \ref{fig:bstar1} and \ref{fig:bstar2} it can be immediately seen that there is a generally good agreement between models and observations: as expected, the rotation rapidly decreases with decreasing surface gravity. 
However, for currently slow-rotating stars, magnetic braking inhibits our ability to access their prior rotational history, and in particular how rapidly they might have been rotating initially.
We emphasize again that this result is only achieved when surface magnetic fields are accounted for in the models; otherwise, in general, only initially slow-rotating (or, in practice, non-rotating) models are consistent with observations of currently slow-rotating stars. 

The median equatorial surface rotational velocity of the sample stars is 78~km~s$^{-1}$ and more than half of the sample stars have a surface rotational velocity less than 100 km s$^{-1}$. It is also apparent that the most massive stars in the sample are all slow rotators ($<$ 50 km s$^{-1}$). A smaller fraction of stars (36.3\%) is located between 100 and 200 ~km~s$^{-1}$, while three single stars and one binary star (7.3\%) have notably high rotational velocities ($>$ 200 ~km~s$^{-1}$). 

While, in general, the agreement between the models and observations is good, a striking discrepancy (in particular, on the left panel of Figure~\ref{fig:bstar2}) is seen in the case of HD\,122451Ab ($\log g = 3.55$, $v_{\rm eq} = 154$ km s$^{-1}$, $B_{\rm p} \approx 230$~G), which is in a triple-system and a known $\beta$ Cep pulsator \citep{pigulski2016}. The orbital period of the two closer components (HD\,122451Aa,b) is nearly one year, on a highly eccentric orbit. It can therefore be speculated that the magnetic star gains angular momentum from the orbital reservoir, and thus rotates more rapidly than expected from a single star undergoing magnetic braking. 

The three most rapidly-rotating ($>$ 200~km~s$^{-1}$) single stars in the sample (HD\,182180, HD\,142184, HD\,345439) require some more attention. These stars have order of 10 kG dipolar magnetic fields, consistent with the mean strength near the ZAMS, where their high log~$g$ values indicate they reside. Their rapid rotation is further consistent with a young age. Importantly, our models predict no such stars should be present except at very young ages, as is indeed the case.

The presence of young and rapidly-rotating magnetic stars is interesting in light of theoretical considerations regarding the ZAMS rotation rates of magnetic stars. The fossil field hypothesis relies on the long-term stability of magnetic fields and assumes that these fields were generated during the star formation process or at the latest during the pre-main sequence evolution of the star \linebreak\citep{donati2009}. In lower-mass Ap and Herbig Ae/Be stars, pre-main sequence magnetic braking is indeed evidenced \citep{stepien2000,alecian2013}. This could also be the case for O and B stars (where the pre-main sequence phase and the magnetic fields are not practically observable due to these stars being embedded in dust). However, the contraction time-scale is shorter than the magnetic braking time-scale, hence a magnetic star can arrive at the ZAMS as a fast rotator. In fact, initially moderate and high rotation rates at the ZAMS are clearly required to explain almost half of the sample of known magnetic B stars. 
Therefore, reconciling the rapidly-rotating stars with our models leads us to conclude that they must initiate their main sequence evolution with nearly critical rotation. If, on the other hand, these stars did not start their main sequence evolution as rapid rotators, consideration of \textit{spin up} mechanisms would be required. 

An intriguing scenario is to explain the observed rapid rotation of some stars via the binary channel, assuming that a companion star may be responsible for either halting magnetic braking or, in fact, spinning up the magnetic star via tidal interactions or mass transfer. Indeed, a significant fraction of OB stars are expected to be in binary systems \citep{demink2014,sana2014}. As of now there is no observational evidence that either of the three rapidly-rotating stars has a close companion star. However, if such a surprising discovery were made, it would provide a natural explanation for their short rotation periods. Since the incidence rate of magnetism in close binary system is very low \citep{alecian2013}, another alternative to the binary channel is that magnetic massive stars may be merger products, thus initially a system of two (pre-) main sequence binary stars \citep{ferrario2009,schneider2016}. This may also result in a rapidly-rotating star, although likely only for a short time-scale. Interestingly, in recent numerical simulations of binary mergers, \cite{schneider2019} obtain a post-merger object with a moderate rotational velocity ($\approx$~200 km\,s$^{-1}$) at the ZAMS.

%
%
\begin{figure*}
\includegraphics[width=9cm]{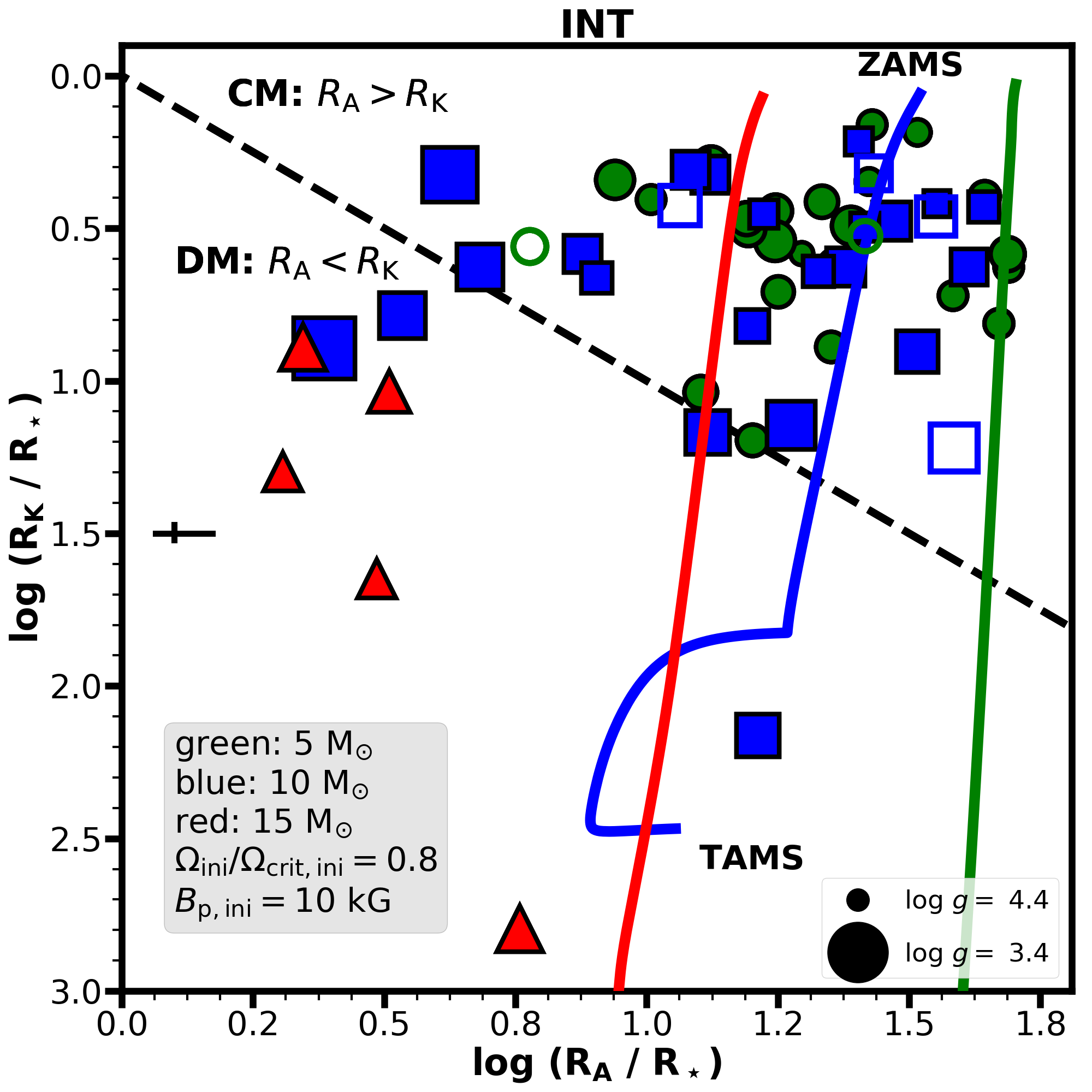}\includegraphics[width=9cm]{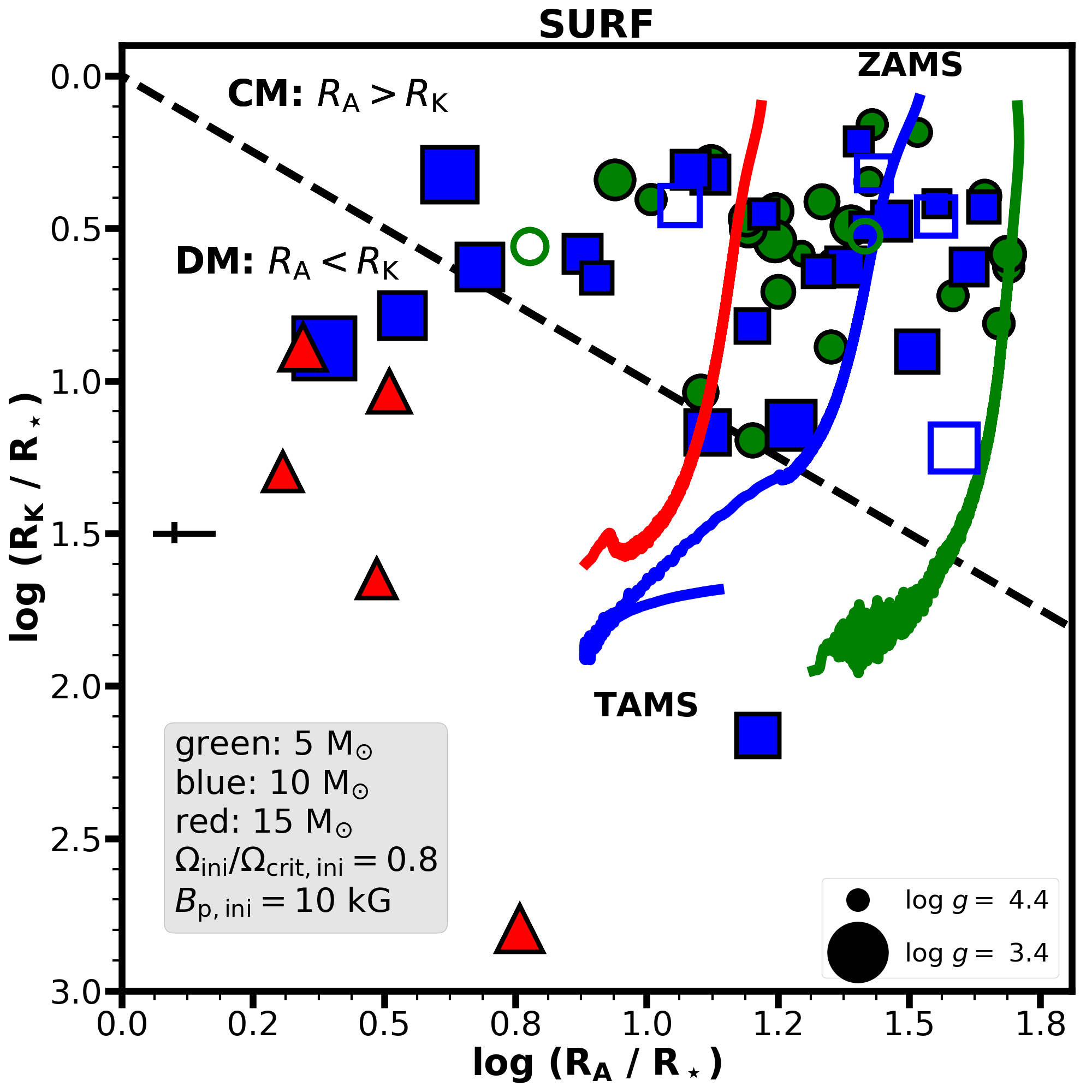}
\caption{Shown is the rotation-confinement diagram, i.e., the Alfv\'en radius vs.\ the Kepler co-rotation radius. The observations are separated into three mass bins: 5 (green circles), 10 (blue squares) and 15 (red triangles) M$_\odot$. Models with $\Omega_{\rm ini}/\Omega_{\rm ini,crit} = 0.8$ and an initial polar field strength of 10 kG are shown. Known short-period binaries are shown with open symbols. The marker size corresponds to the logarithmic surface gravity. The mean errorbars are shown to the centre left. DM and CM indicates dynamical and centrifugal magnetospheres, respectively. The model evolution begins at the top right corner of the diagrams. \textit{Left:} internal magnetic braking. \textit{Right:} surface magnetic braking.  }\label{fig:bstar4}
\end{figure*}
%
%

\subsubsection{Magnetic evolution of known magnetic B-type stars}

Figures \ref{fig:bstar1} and \ref{fig:bstar2} show that, in general, models with a large range of initial magnetic field strength are in agreement with observations. 

\cite{2019MNRAS.490..274S} showed that no clear trend could be identified between $B_{\rm p}$ and $M_\star$ (their Figure 9), and they argued that the magnetic field strength may decrease more rapidly than expected from magnetic flux conservation (see also \citealt{landstreet2007, landstreet2008,fossati2016}). Indeed, the weaker magnetic fields of the more massive B-type stars are a consequence of those stars being more evolved.

A few presumably evolved stars with weak magnetic fields are not matched by the computed models (see Figures~\ref{fig:bstar1} and \ref{fig:bstar2}). Since the observed magnetic field strengths span a large range, it may be possible that those stars with low values of $\log g$ and $B_{\rm p}$ simply initiated their evolution with weaker fields than considered in our models ($< 1$ kG). However, their progenitors (with high $\log g$ and low $B_{\rm p}$) remain undetected. Another piece in this puzzle is the lack of descendants with strong fields (which are easier to detect) but lower surface gravity. The absence of such stars is consistent with previous suggestions that the magnetic fields of OB stars are subject to a flux decay mechanism, in contrast to the evolution of the magnetic fields of A and Ap stars which are consistent with flux conservation \linebreak[4]\citep{landstreet2007,neiner2017,sikora2019}. The flux decay rate could, in principle, be inferred empirically by matching the observed relationship of log~$g$ and $B_{\rm p}$. If a field decay in the computed models needed to be accounted for, then the tracks with given initial field strength would show lower magnetic field for a given surface gravity. On the other hand, this weakening may affect the rotational evolution, which will need to be studied in detail with models accounting for different field evolution scenarios.

\subsection{Evolution of rotation and confinement} 

The rotation-confinement diagram (Figure \ref{fig:bstar4}, see also Figure 3 of \citealt{petit2013}, Figure 7 of Paper I, and Figure 8 of \citealt{2019MNRAS.490..274S}) shows the relative importance of magnetic wind confinement and centrifugal support in terms of the Alfv\'en radius and the Keplerian co-rotation radius (Equations \ref{eq:alf} and \ref{eq:rk}). The overwhelming majority of the observed B-type stars are in the centrifugal magnetosphere (CM) regime (see Section~\ref{sec:two}). The significance of the magnetospheric characterisation is related to the age determination of observed stars, as stars with CMs are typically expected to be young, whereas stars with dynamical magnetospheres (DMs) are expected to be evolved (\citealt{petit2013}, Paper I, \citealt{2019MNRAS.490..274S}). 

The evolutionary models shown in Figure \ref{fig:bstar4} have an initial field strength of 10 kG. \cite{2019MNRAS.490..274S} found that all early B-stars in the final third of their MS lifetimes have DMs. The absence of evolved stars with CMs is consistent with the typical ZAMS surface magnetic field strength being higher than 1 kG.

The stars in the 15\,M$_\odot$ mass bin from the \cite{shultz2018} sample have a DM, therefore they are expected to be evolved. However, they are not matched with the computed models on the $R_{\rm K} - R_{\rm A}$ plane. A possible explanation is that the magnetic field strength weakens more rapidly than expected from magnetic flux conservation (see previous section), in which case the decrease in $R_{\rm A}$ is more rapid, thus our tracks on the diagram would indicate lower $R_{\rm A}$ values for given $R_{\rm K}$. This correction could reconcile the discrepancy. Another possibility is if magnetic braking was much more efficient on the pre-main sequence for more massive stars, so they arrive at the ZAMS already as slow rotators. It could also be that at least some magnetic stars follow an unusual trend on the HRD. In particular, HD 149438 ($\tau$ Sco) may be a blue straggler and is a promising candidate for a merger product \citep{schneider2016,schneider2019}. Furthermore, $\tau$ Sco is also a significant example of a magnetic star that may present an apparently younger age than the age of the cluster that it belongs to \citep{nieva2014,schneider2016}. If this is a general trend, this would mean that for the sample stars the cluster ages may overestimate their actual age.

The stars in the 10\,M$_\odot$ bin show a scatter, however, leading to similar findings as in the case of the 15 M$_\odot$ stars. The stars in the 5\,M$_\odot$ bin are all located in the CM regime (except one star). This is consistent with their rotational and magnetic field evolution as both surface rotation and magnetic field strength show high values (Figures \ref{fig:bstar1} and \ref{fig:bstar4}). This suggests that these stars are mostly in their early evolutionary stage, in agreement with the findings of \cite{2019MNRAS.490..274S}. The INT models predict 5-10 M$_\odot$ stars in the lower part of the diagram (large $R_{\rm A}$, small $R_{\rm K}$, see Figure~\ref{fig:bstar4}), while the SURF models do not. The latter case seems more consistent with the lack of such observations.

All known close binary stars are in the CM regime. Indeed, HD\,36485, HD\,37017, and HD\,156324 are in the extreme CM regime and display H$\alpha$ emission, which correlates strongly with strong magnetic confinement, rapid rotation, and young stellar age. However, to what degree this is a consequence of tidal acceleration is not clear. The three binary stars showing H$\alpha$-emission are all very young, and their CMs are no larger than single H$\alpha$-bright stars of comparable ages \citep{2019MNRAS.490..274S}. The two close binary systems without H$\alpha$ emission, HD\,136504 (in which system both components are magnetic, \citealt{shultz2015}) and HD\,149277, are middle-aged and have CMs that are somewhat smaller than the average of their contemporaries.

\subsection{Remarks from the model validation from the comparison with known magnetic B-type stars}

The purpose of this initial comparison between state-of-the-art models and a sample of observed magnetic B-type stars was to validate the models that  contain a prescription of surface fossil magnetic fields. We found that in general the models are in good agreement with the observations when considering their rotational and magnetic evolution alone. 

Three observed (presumably single) stars and some stars with moderate rotation (100-200 km s$^{-1}$) are only compatible with initially (at least) moderately or rapidly-rotating models. The rest of the sample (consisting of mostly older, slowly-rotating stars) leads to a degenerate solution to determine their initial rotation rates. Observations require models with a \textit{range} of initial rotational velocities. This has an impact on star formation theory, which will have to account for both the fast and slow ZAMS rotation of massive stars with surface fossil magnetic fields. 

There is no striking discrepancy between the models and observations when looking at the magnetic field evolution alone. However, from the rotational and magnetospheric diagnostics, one could argue that a systematic discrepancy is introduced by adopting magnetic flux conservation as the modelled field evolution. Indeed, a systematic shift of the models is expected if magnetic flux decay (i.e., a more rapid decline than expected from the $\propto R_\star^{-2}$ dependence alone) can be quantified and hence implemented in our models. This could potentially be the key to reconciling the evolution seen on the $R_{\rm K} - R_{\rm A}$ plane.

%
%

\section{Conclusions} \label{sec:con}

In this work, we described the implementation of magnetic braking applicable for hot, massive stars in the \textsc{mesa} software instrument, and studied the rotational evolution of the models. We provide the scientific community with this additional \textsc{mesa} module that contains a realistic and simple prescription of surface fossil magnetic fields in stellar evolution codes (see also \citealt{keszthelyi2017a,petit2017,georgy2017,keszthelyi2019}). We emphasize; however, that this implementation needs to be improved to consider additional components for a more comprehensive picture. 

Presently, there exists no verified formalism that could treat internal angular momentum transport by a fossil field in massive stars; although some approaches have been explored - and many more in the context of dynamo-generated fields. In principle, the contribution from Poynting stresses could be considered in a similar manner; however, the picture is likely to be far more complex and first the interaction with hydrodynamical instabilities would need to be established. Empirical evidence may be collected soon with the advent of magneto-asteroseismology \citep{bram2018,prat2019}, and it will be invaluable to guide evolutionary modelling. One of the key uncertainties remains the internal effects of the fossil field and the extent to which they are anchored inside the star. We hope that the parameterization developed in this work will become a convenient tool to adjust the number of torqued layers in the stellar model. To improve on the implementation presented in this work, one may consider coupling it with other approaches that focused on different aspects. For example, the interaction of fossil fields and convection \citep{feat2009,petermann2015,macdonald2019}, structural changes by the fossil field \citep{mathis2005,duez2010,prat2019}, the magneto-rotational instability \citep{wheeler2015,quentin2018}, and appropriate angular momentum transport \citep{spruit2002,maeder2005,heger2005,denissenkov2007,rudiger2015,kissin2018,fuller2019,schneider2019} would be required to establish all-together a more coherent picture regarding the effects of surface fossil magnetic fields. 

The models presented in this study are one-dimensional. As such, they naturally ignore any variations in co-latitude. Based on current two-dimensional MHD simulations, it is not clear how well the assumed angular momentum loss via magnetic braking (Equation \ref{eq:eq1}) applies in more realistic three-dimensional situations where the magnetic field is tilted with respect to the rotation axis. It is possible that the tilt would only lead to minor corrections in the formalism and our assumptions here are still appropriate.

%
%

We found that the time evolution of magnetic braking is an essential component since it is rapidly weakened by stellar feedback effects - primarily via decreasing the surface rotational velocity. We demonstrated that internal magnetic braking can greatly deplete the total angular momentum reservoir, whereas surface magnetic braking allows the star to maintain most of its total angular momentum.

%
%

With the inclusion of surface fossil magnetic fields in stellar evolution models, we identified that initially fast-rotating magnetic models undergo an early blueward evolution. Models with given surface fossil magnetic fields and initial mass but different initial rotation rates, merge to a common track, leading to a degeneracy to disentangle between their initial rotation rates. Therefore, further considerations are required before gyrochronology could be applied to magnetic massive stars. Moreover, we showed that initially fast-rotating magnetic stellar models evolve quite differently than non-magnetic ones, causing a significant discrepancy in the derived stellar ages, amongst others.

%
%

Comparing our models with observations from \cite{shultz2018}, we found that most likely a range of initial rotation rates is required to explain both slowly and rapidly-rotating young magnetic B-type stars. This has potential consequences to explain the formation of magnetic massive stars. Furthermore, this may help to shed more light on fossil field evolution: if magnetic massive stars arrive at the ZAMS with close to critical rotation, then pre-main sequence magnetic braking is either inefficient or absent.

%
%

Models with surface and internal magnetic braking were shown to be non-interchangeable: they produce different results. Both our SURF and INT models might be compatible with the comprehensive sample of known magnetic B-type stars of \cite{shultz2018}. Further studies should be conducted to evaluate these scenarios.

\section*{Acknowledgements}

We thank the anonymous referee for providing us with a constructive and helpful report, which has led to improving the paper.
We appreciate discussions with K. Augustson and P. Marchant. We are grateful for B. Paxton and the \textsc{mesa} developers for making their code publicly available. 
G.M. and C.G. acknowledge support from the Swiss National Science Foundation (project number 200020-172505).
M.E.S. acknowledges support from the Annie Jump Cannon Fellowship, which supported by the University of Delaware and is endowed by the Mount Cuba Astronomical Observatory.
A.uD. acknowledges support from NASA through Chandra Award number TM7-18001X issued by the Chandra X-ray Observatory Center, which is operated by the Smithsonian Astrophysical Observatory for and on behalf of NASA under contract NAS8-03060. 
R.H.D.T. acknowledges support from National Science Foundation grants ACI-1663696 and AST-1716436.
G.A.W. acknowledge support in the form of a Discovery Grant from the Natural Science and Engineering Research Council (NSERC) of Canada. 
V.P. acknowledges support provided by 
(i) the National Aeronautics and Space Administration through Chandra Award Number GO3-14017A issued by the Chandra X-ray Observatory Center, which is operated by the Smithsonian Astrophysical Observatory for and on behalf of the National Aeronautics Space Administration under contract NAS8-03060, and
(ii) program HST-GO-13734.011-A that was provided by NASA through a grant from the Space Telescope Science Institute, which is operated by the Association of Universities for Research in Astronomy, Inc., under NASA contract NAS 5-26555.
A.D.U acknowledges support from NSERC. 
%
%

\bibliographystyle{mnras}
\bibliography{ref}


\appendix

%
\section{Implementation in GENEC}\label{sec:genec}

The structure of the star in \textsc{genec} is slightly different to the \textsc{mesa} one. The main differences are the following:
\begin{itemize}
\item In \textsc{genec}, the structure equations, chemical composition evolution equations, and rotation equations are decoupled and solved one after the other iteratively up to a converged structure is reached at a given time step. In \textsc{mesa}, the whole set of coupled equations are solved at once.
\item In \textsc{genec}, the star is divided into three different regions \citep[see also][]{Kippenhahn1990a}: the interior, going from the centre of the star to a given `fitting point', located at a mass coordinate $M_\text{fit}$ inside the star; the envelope, going from $M_\text{fit}$ to the point where an optical depth of $\tau = 2/3$ is reached; the atmosphere, providing the boundary conditions. In the interior, the whole set of equations is solved. In the envelope, the energy generation rate is assumed to be zero, the chemical composition constant, and the rotation uniform \citep[see][for more details]{Maeder2009a}.
\item The time evolution of the angular momentum transport in \textsc{genec} is computed by solving the following advecto-diffusive equation \citep{maeder1998}:
\begin{equation}
\rho\frac{\text{d}}{\text{d}t}\left(r^2\Omega\right)_{M_r} = \frac{1}{5r^2}\frac{\partial}{\partial r}\left(\rho r^4\Omega U(r)\right) + \frac{1}{r^2}\frac{\partial}{\partial}\left(\rho Dr^4\frac{\partial\Omega}{\partial r}\right),\label{eq:eq_rotGE}
\end{equation}
with $r$ the radius, $\rho$ the density, $\Omega$ the angular velocity, $U(r)$ the radial component of the meridional circulation, and $D$ a diffusion coefficient accounting for various instabilities. This equation is solved following an operator-splitting approach, where the advective part and the diffusive part are solved alternatively every two time steps. \textsc{mesa} uses a fully diffusive approximation, which is practically identical to Equation \ref{eq:eq_rotGE} without the advective term (the first term on the r.h.s.). 
\end{itemize}
To account for magnetic braking in \textsc{genec}, an additional torque is applied as a boundary condition to Equation \ref{eq:eq_rotGE} \citep[see][for a detailed discussion]{eggenberger2008}. In this framework, the fitting mass $M_\text{fit}$ plays the same role as the mass $M_t$ described in the discussion about the implementation in \textsc{mesa}. However, instead of assuming that the same fraction of the angular momentum is lost in each layer, this zone is assumed to rotate as a solid body, and the angular momentum is removed in the entire region keeping solid-body rotation.

%
%
\section{Convergence test between the SURF and 
INT cases}\label{sec:a1}
\begin{figure*}
\includegraphics[width=6cm]{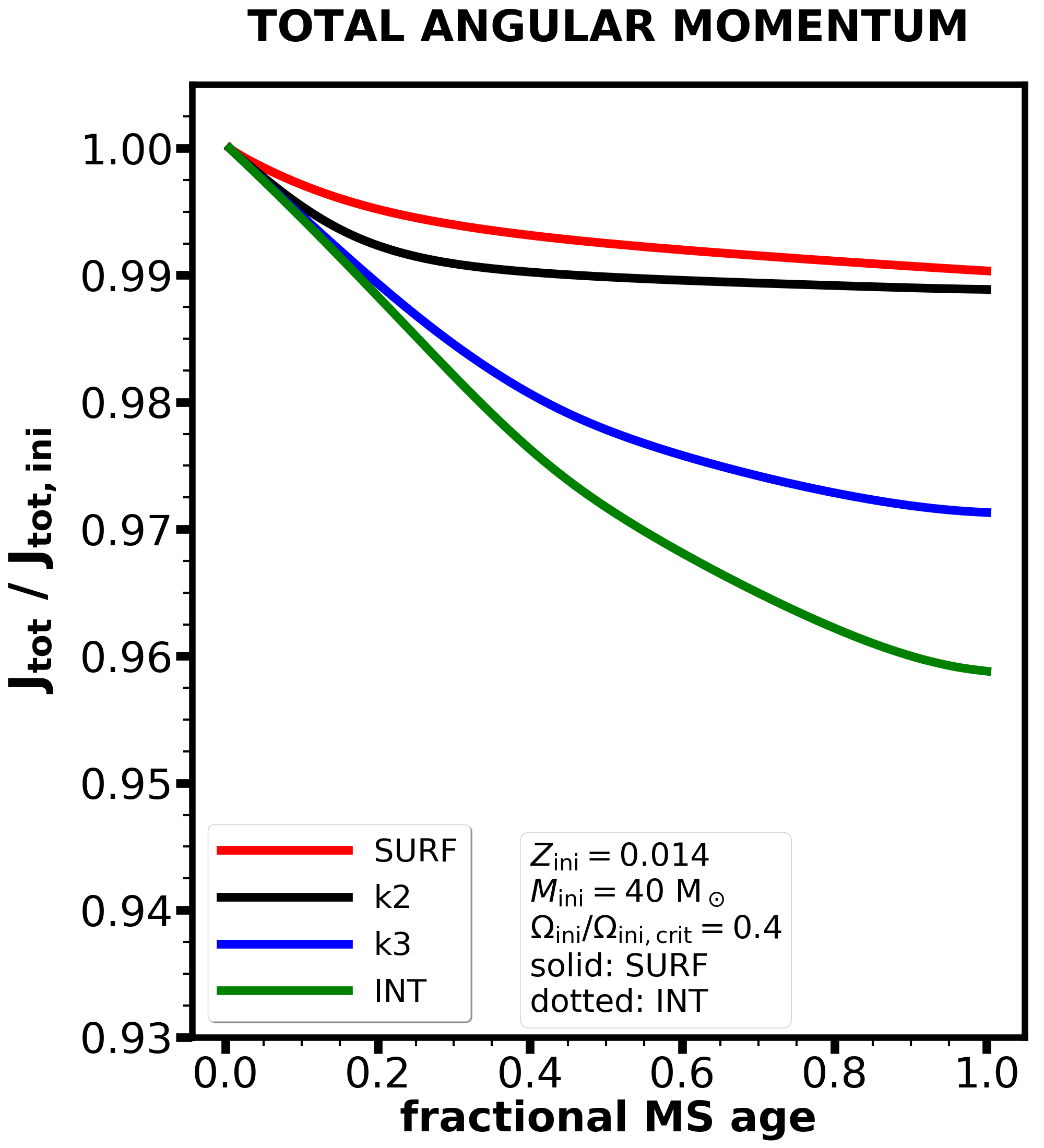}\includegraphics[width=6cm]{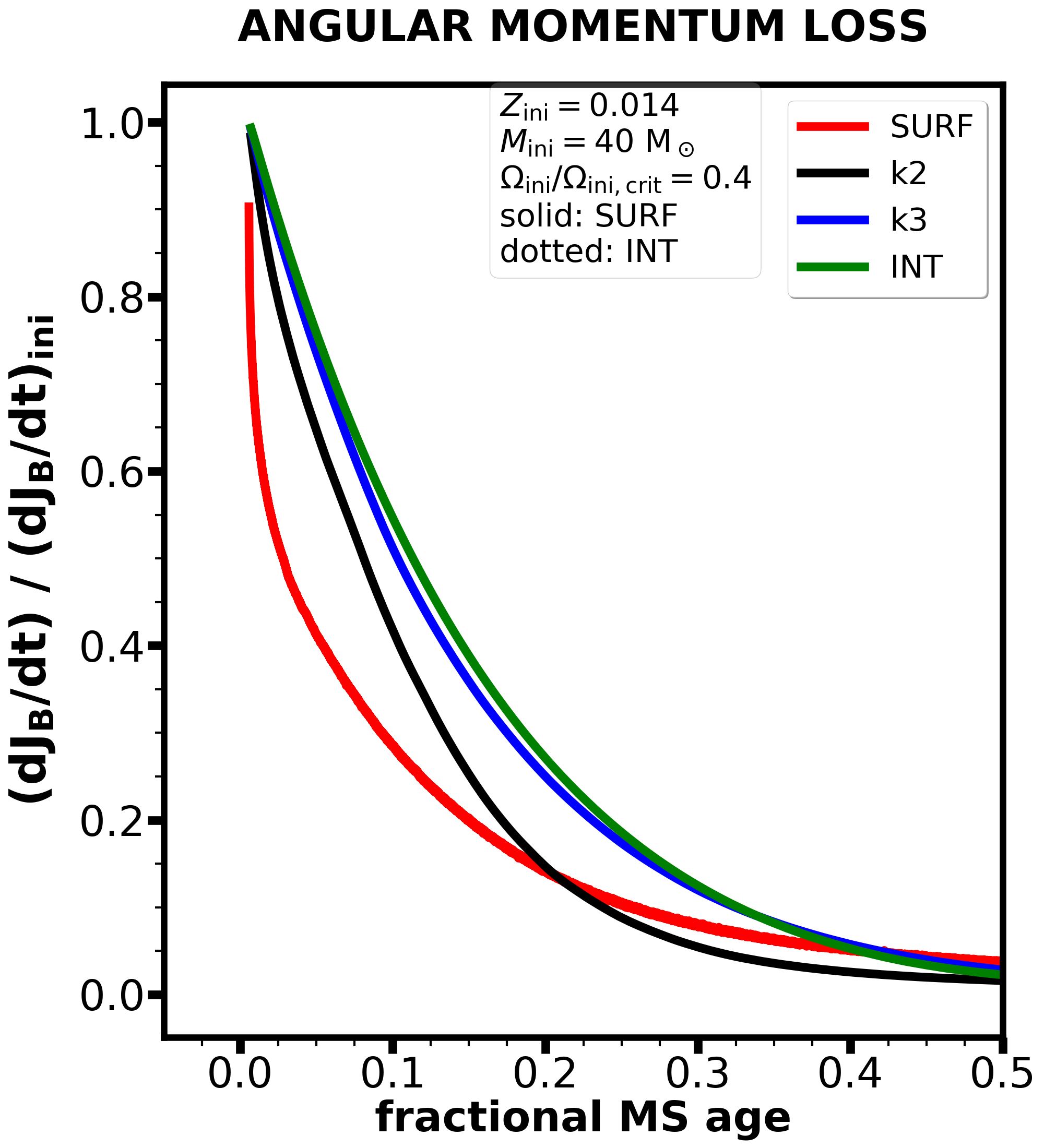}\includegraphics[width=6cm]{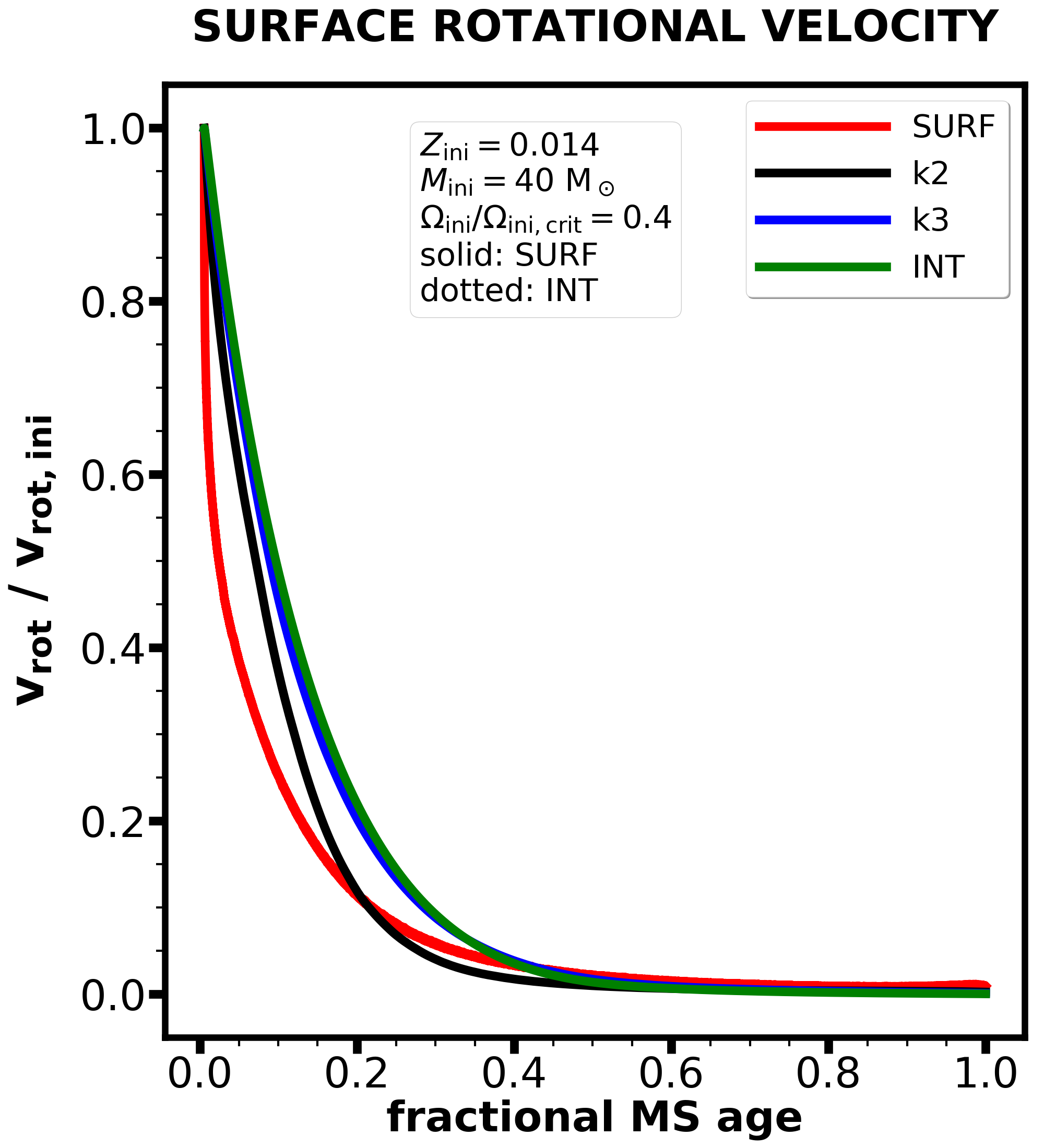}
\caption{Evolution of the total angular momentum, the angular momentum loss rate driven by the magnetic field and the gas, and surface rotational velocity compared to their initial values. k2 and k3 denote models where the number of torqued layers are doubled and tripled compared to the SURF case.}\label{fig:app1}
\end{figure*}
In the following parameter test, the same input is used as described for our `default' ($M_{\rm \star, ini}=$40~M$_\odot$, $\Omega_{\rm ini}/\Omega_{\rm ini, crit} = 0.4$, $B_{\rm p, ini} = 5$ kG) model. For the four models presented in Figure~\ref{fig:app1}, the number of torqued layers vary as follows: SURF denotes the case when $x = k\_const\_mass$ as described in Section~\ref{sec:impl}, k2 denotes $x = 2 \cdot k\_const\_mass$, k3 denotes $x = 3 \cdot k\_const\_mass$, and in the INT case the angular momentum loss is distributed to every layer of the star. Since the number of layers are increased, in practice, this corresponds to distributing the angular momentum loss to larger and larger mass fractions of the star.
Thus we verify that the SURF case tends toward the INT case when the number of torqued layers are increased.

\section{Tables}
%
%
%
\begin{table*}
\caption{15 M$_\odot$ model comparison at half-way through their main sequence life-times from Figure \ref{fig:mhh} (indicated with purple marker).}
\label{tab:t2}
\centering
\begin{tabular}{llc|cccccccccc}  
\hline  \hline
model & braking & $\frac{\Omega_{\rm ini}}{\Omega_{\rm crit, ini}}$ & $T_{\rm eff}$ & log $L/L_{\odot}$ & $M_\star$& log $g$ & $v_{\rm rot}$ & log $J_{\rm tot}$  & age \\
 &  scheme & & [kK]  & &  [M$_{\odot}$] &[cm s$^{-2}$]  &[km s$^{-1}$] &[g cm$^2$ s$^{-1}$] & [Myr] & \\
\hline \hline
MAG & INT &  0.2 & 29.541 & 4.435 & 14.994 & 4.014 & 38.946 & 51.528 & 6.532 \\
MAG & INT &  0.5 & 29.508 & 4.433 & 14.994 & 4.014 & 82.860 & 51.838 & 6.583  \\
MAG & INT &  0.8 & 29.605 & 4.433 & 14.994 & 4.020 & 90.946 &  51.874 & 6.683  \\
\hline 
MAG & SURF &  0.2 & 29.561 &  4.435 &  14.994 & 4.015 &  20.364 &  51.825 & 6.551  \\
MAG & SURF &  0.5 & 29.715 &  4.438 &  14.994 & 4.021 &  49.061 &  52.012 & 6.769 \\
MAG & SURF &  0.8 & 29.923 &  4.439 &  14.994 & 4.032 &  51.828 &  52.048 & 6.944 \\
\hline
NOMAG & - &  0.8 & 26.983 &  4.404 &  14.888 & 3.885 & 514.516 &  52.540 & 7.363  \\
\end{tabular} 
\end{table*}
%
%
%
%
\begin{table*} 
\caption{15 M$_\odot$ model comparison at the TAMS from Figure \ref{fig:mhh}.}
\label{tab:t3}
\centering
\begin{tabular}{llcccccccccccc}   
\hline \hline
model & braking & $\frac{\Omega_{\rm ini}}{\Omega_{\rm crit, ini}}$ & $T_{\rm eff}$ & log $L/L_{\odot}$ & $M_\star$& log $g$ & $v_{\rm rot}$ & log $J_{\rm tot}$  & age \\
 &  scheme & & [kK]  & &  [M$_{\odot}$] &[cm s$^{-2}$]  &[km s$^{-1}$] &[g cm$^2$ s$^{-1}$] & [Myr] & \\
\hline \hline
MAG & INT &  0.2 & 24.007 &  4.731 &  14.982 & 3.357 &  6.134 &  50.904 &  13.059 \\
MAG & INT &  0.5 & 24.040 &  4.731 &  14.982 & 3.360 &  10.488 &  51.080 &  13.163 \\
MAG & INT &  0.8 &24.122 &   4.736 &  14.983 & 3.361 &  7.433 &  50.945 & 13.362 \\
\hline 
MAG & SURF &  0.2 & 24.089 &  4.731 &  14.982 &3.363 &  8.737 &  51.683 &  13.100  \\
MAG & SURF &  0.5 & 24.124 &  4.749 &  14.981 &3.348 &  10.381 &  51.816 & 13.538 \\
MAG & SURF &  0.8 & 24.258 &  4.761 &  14.980 &3.345&  8.565 & 51.832 &  13.887 \\
\hline
NOMAG & - &  0.8 & 21.732 & 4.732 & 14.542 & 3.170 &  306.788 & 52.305 & 14.726 \\
\end{tabular} 
\end{table*}
%



\bsp	
\label{lastpage}
\end{document}